\documentclass[aps,prd,12point,twocolumn,nofootinbib,showpacs,superscriptaddress]{revtex4-2}

\usepackage{graphicx}
\usepackage{dcolumn}
\usepackage{tabu}
\usepackage{amsmath,amssymb}
\usepackage{float}
\usepackage{natbib}
\usepackage{balance}
\usepackage[normalem]{ulem}
\usepackage{braket}
\usepackage{upgreek}
\usepackage{mathrsfs}
\usepackage{bm}
\usepackage{lipsum}
\usepackage{comment}
\usepackage{xfrac}
\usepackage[dvipsnames]{xcolor}
\usepackage[title]{appendix}
\usepackage{txfonts}
\usepackage{xcolor}
\usepackage{slashed}
\usepackage[breaklinks=true,pdfborder={0 0 0},bookmarksopen]{hyperref}
\hypersetup{
	colorlinks   = true, 
	urlcolor     = magenta, 
	linkcolor    = blue, 
	citecolor   = Red 
}

\begin{document}

\title{Spin precession and neutrino helicity flip in various spacetimes}

\author{R. Saadati}
\email{rsaadati@ubishops.ca}
\affiliation{Department of Physics \& Astronomy,
Bishop's University,\\
2600 College Street, Sherbrooke, QC, J1M 1Z7, Canada}

\author{F. Hammad}
\email{fhammad@ubishops.ca}
\affiliation{Department of Physics \& Astronomy,
Bishop's University,\\
2600 College Street, Sherbrooke, QC, J1M 1Z7, Canada}
\affiliation{Physics Department, Champlain College-Lennoxville,\\
2580 College Street, Sherbrooke, QC, J1M 0C8, Canada}

\author{S. Novoa-Cattivelli}
\email{snovoa23@ubishops.ca}
\affiliation{Department of Physics \& Astronomy,
Bishop's University,\\
2600 College Street, Sherbrooke, QC, J1M 1Z7, Canada}

\author{M. Simard}
\email{msimard23@ubishops.ca}
\affiliation{Department of Physics \& Astronomy,
Bishop's University,\\
2600 College Street, Sherbrooke, QC, J1M 1Z7, Canada}

\author{N. Fleury}
\email{nfleury22@ubishops.ca}
\affiliation{Department of Physics \& Astronomy,
Bishop's University,\\
2600 College Street, Sherbrooke, QC, J1M 1Z7, Canada}


\begin{abstract}
We systematically study spin precession of neutral particles freely moving in spacetime. We first derive the formula describing spin precession within a general stationary and axisymmetric spacetime. We then apply our formula to study spin precession of neutral spinning particles moving within various spacetimes that are among the most ubiquitous ones found in the literature. Our results are then used to extract the helicity flip probability for neutrinos propagating in each one of those selected spacetimes. It is found that low-energy neutrinos acquire a spin-flip probability that is as large as unity in all the spacetimes considered here. The remarkable result, however, is that while spin-flip probability for high-energy neutrinos remains insignificant in most of these spacetimes, some of them do allow spin-flip probability to reach unity even for high-energy neutrinos.
\end{abstract}

\maketitle
\section{Introduction}\label{sec:Intro}
Besides the three classical tests of general relativity (namely, the deflection of light by the Sun, the gravitational red-shift of light and the anomalous advance of Mercury’s perihelion), general relativity predicts the precession (that was later experimentally confirmed \cite{Will}) of the spin of massive bodies. However, although spin precession and the dynamics of spinning bodies in general relativity have been derived for classical bodies like a gyroscope \cite{DeSitter,LensThirring} when considered to be point-like and then for classical bodies whose extension is taken into account \cite{Mathisson,Papapetrou,Dixon1964,Dixon}, the possibility of applying those formulas to spinning \textit{quantum} particles has been seriously considered in the literature very early on as well \cite{Papini1991,Casini1994,Papini2004}. 

Yet, another application that seems very promising for experimental investigations of gravitational physics using current technology is the study of neutrino helicity flip under the effect of gravity \cite{Dvornikov2006,Dvornikov2013,Alavi,Dvornikov2019,Dvornikov2020a,Dvornikov2020b,PRD2021}. Indeed, being always emitted by sources and detected by detectors as left-handed particles, a flip of their helicity would make neutrinos escape detection altogether. On the other hand, neutrinos are abundant particles that  travel tremendous interstellar distances without being absorbed. Therefore, a deficit in the number of detected neutrinos that would be caused by a gravitationally induced helicity flip would convey valuable information about the gravitational field through which the particles have traveled before reaching the detector. Given the insignificance of the gravitationally induced neutrino flavor oscillations compared to matter induced neutrino flavor oscillations \cite{NWPG,ConfNO}, the gravitationally induced neutrino helicity flip might offer a serious neutrino-based alternative for probing gravity.

Our first aim in this paper is to systematically study spin precession of neutral particles freely falling in a gravitational field by deriving a formula for general stationary and axisymmetric spacetimes. The formula can then be applied to various ubiquitous spacetimes found in the literature. The results that would come out of this first aim of the paper will be general as they are not limited to any specific spinning particle. Our second aim is to apply those general results to the case of neutrinos and derive the helicity flip probability for neutrinos traveling in those various spacetimes.

To achieve our two aims, we structured our paper as follows. In Sec.\,\ref{Sec:Precession}, we give a very brief summary of the formulas and tools used to describe spin precession in curved spacetime. In Sec.\,\ref{Sec:InStationaryAST}, we use our general formula from Sec.\,\ref{Sec:Precession} to derive the angular velocity of spin precession of neutral particles freely moving within stationary and axisymmetric spacetimes. An application to thirteen different spacetimes will then be made in that section. In Sec.\,\ref{Sec:HelicityFlip}, we rely on those results to derive the neutrino helicity flip probability formula for neutrinos propagating in each one of those specific spacetimes. We summarize and discuss the relevance of our results to observational astrophysics and cosmology in the brief Sec.\,\ref{Sec:Conclusion}.

\section{Spin precession in curved spacetime}\label{Sec:Precession}
Spin precession of macroscopic bodies, like a gyroscope or a small but extended particle, is dealt with using the Mathisson-Papapetrou-Dixon (MPD) equations \cite{Mathisson,Papapetrou,Dixon} (see also \cite{BakerReview,Deriglazov} or Ref.\,\cite{MTW} for a textbook presentation.) The MPD equations describing spin dynamics are a set of nonlinear equations written in a covariant form for the spin tensor $S^{\mu\nu}$. The latter is related to the spin four-vector $S^\mu$ of the particle by $S^\mu=-\frac{1}{2}\epsilon^\mu_{\;\,\nu\rho\lambda}p^\nu S^{\rho\lambda}$, where $p^\mu=mu^\mu$ is the four-momentum of the particle of mass $m$ and of four-velocity $u^\mu$, and $\epsilon_{\mu\nu\rho\lambda}$ is the totally antisymmetric Levi-Civita tensor\footnote{We shall work with the natural units $G=c=\hbar$ and we adopt the metric signature $(-,+,+,+)$.}. At the zeroth order in spin, a case which is amply sufficient to our case, the MPD equations reduce to the simple equation ${\rm D}S^\mu/{\rm d}\tau=0$ \cite{MTW}, where ${\rm D}$ is the covariant differentiation operator, and $\tau$ is the proper time of the particle. To cast the equation into the comoving frame of the particle, one uses the spacetime comoving vierbeins $e_\mu^{\hat a}$, defined by  $\eta_{{\hat a}{\hat b}}e_\mu^{\hat a}e_\nu^{\hat b}=g_{\mu\nu}$ for any spacetime metric $g_{\mu\nu}$, with the corresponding comoving inverse vierbeins $e^\mu_{\hat a}$ defined by $e^{\hat a}_\mu e^\mu_{\hat b}=\delta^{\hat a}_{\hat b}$. Here, $\eta_{{\hat a}{\hat b}}$ is the Minkowski metric and $\delta_{\hat b}^{\hat a}$ is the Kronecker delta symbol\footnote{The Greek letters are used to denote spacetime indices, whereas the first Latin letters $(a,b)$ are used to denote tangent-space indices. The Latin letters $(i,j)$ are used to denote indices of the three-dimensional space. Hats over Latin letters denote indices in the comoving frame.}.

In the comoving frame of the particle, the proper-time evolution of the spin three-vector $\textbf{S}$ (having components $S^{\hat i}$) obeys the following first-order differential equation:
\begin{equation}\label{S^a Precession}
\frac{{\rm d}S^{\hat i}}{{\rm d}\tau}=-S_{\hat a}u^\mu\omega_\mu^{\,{\hat i}\hat{a}}.
\end{equation}
Here, $\omega_\mu^{\,{\hat a}{\hat b}}$ is the spin connection, defined in terms of the comoving vierbeins and the Christoffel symbols $\Gamma_{\mu\nu}^\lambda$  corresponding to the spacetime metric by 
\begin{equation}\label{SpinConnectionDef}
\omega_\mu^{\,\,{\hat a}{\hat b}}=-e^{\nu{\hat  b}}\partial_\mu e^{\hat a}_\nu+e^{\nu {\hat b}}\Gamma_{\mu\nu}^\lambda e^{{\hat a}}_\lambda.
\end{equation}
The angular velocity three-vector $\boldsymbol{\Omega}$ of the spin precession can be immediately read off from the dynamical equation (\ref{S^a Precession}) as the latter is of the form ${\rm d}\textbf{S}/{\rm d}\tau={\boldsymbol{\Omega}}\times\textbf{S}$. Indeed, the three components of the angular velocity then read \begin{equation}\label{OmegaVector}
    \Omega_{\hat i}=\tfrac{1}{2}\epsilon_{{\hat i}{\hat j}{\hat k}}u^\mu\omega_{\mu}^{\,{\hat j}{\hat k}},
\end{equation}
where $\epsilon_{{\hat i}{\hat j}{\hat k}}$ is the totally antisymmetric Levi-Civita symbol, such that $\epsilon_{\hat{1}\hat{2}\hat{3}}=1$. Expression (\ref{OmegaVector}) is valid for any spacetime metric and for any four-velocity $u^\mu$ of the particle. In this paper we are interested in spin precession of particles that are freely propagating in the presence of various gravitational sources that generate various spacetimes. In the next section we shall derive the explicit expression this angular velocity vector takes for arbitrary stationary and axisymmetric metrics before applying the result to specific metrics.

\section{Spin precession within stationary and axisymmetric spacetimes}\label{Sec:InStationaryAST}

Let us consider a general stationary and axisymmetric spacetime metric that has the following form in the spherical coordinates $(t,r,\theta,\phi)$:
\begin{equation}\label{GeneralMetric}
    {\rm d}s^2=-\Lambda\,{\rm d}t^2+\mathcal{R}\,{\rm d}r^2+\Theta\,{\rm d}^2\theta+\Phi\,{\rm d}\phi^2-2\Psi\,{\rm d}\phi\,{\rm d}t.
\end{equation}
The five nonvanishing metric components of this spacetime are all time-independent and depend, \textit{a priori}, on both polar coordinates $r$ and $\theta$.The spacetime metric (\ref{GeneralMetric}) has two Killing vector fields, a timelike Killing vector field that reads $K^\mu=(1,0,0,0)$, and a spacelike Killing vector field that reads $R^\mu=(0,0,0,1)$. The timelike vector field $K^\mu$ yields the conserved energy $\varepsilon=-g_{0\mu} u^\mu$ per unit mass of the particle, and the spacelike vector field $R^\mu$ yields the conserved orbital angular momentum $\ell=g_{3\mu}u^\mu$ per unit mass of the particle. Combining these two identities with the normalization $g_{\mu\nu}u^\mu u^\nu=-1$, corresponding to time-like geodesics of freely falling massive particles, we extract the following general expression for the four-velocity of a particle moving along a plane that is parallel to the equatorial plane of the metric (\ref{GeneralMetric}): 
\begin{equation}\label{General4Velocity}
u^\mu=\frac{1}{\gamma}\left(\alpha,\pm\beta,0,\lambda\right), 
\end{equation}
where
\begin{align}\label{alphabetagamma}
&\gamma=\Psi^2+\Lambda\Phi,\qquad \alpha =\varepsilon\Phi-\ell\Psi,\qquad \lambda=\varepsilon\Psi+\ell\Lambda,\nonumber\\ &\beta=\left[\frac{\alpha\left(\Lambda\alpha+\Psi\lambda\right)+\lambda\left(\Psi\alpha-\Phi\lambda\right)-\gamma^2}{\mathcal{R}}\right]^{\frac{1}{2}}.
\end{align}
The ($+$) sign in Eq.\,(\ref{General4Velocity}) corresponds to a particle moving towards the gravitational source, whereas the ($-$) sign corresponds to a particle moving away from the source. The comoving vierbeins, the Christoffel symbols, and the components of the spin connection corresponding to this metric, as well as the four-velocity, are all displayed in Appendix \ref{Sec:App}. From those expressions, we derive the following general nonvanishing components of the angular velocity vector $\boldsymbol{\Omega}$ of spin precession:
 
\begin{align}\label{GeneralAngularVelocity}
    \Omega_{\hat{1}}&= \pm \frac{1}{2 \gamma  \sqrt{\Delta  \Theta }}\Big\{ (\alpha  \Lambda +\lambda  \Psi ) \Big[\Psi_{,\theta}  \left(\alpha ^2 \Lambda +\lambda ^2 \Phi \right)\nonumber\\
    &\quad+\alpha  \Lambda_{,\theta} (\lambda  \Phi -\alpha  \Psi )-\lambda  \Phi_{,\theta}  (\alpha  \Lambda +\lambda  \Psi )\Big]\nonumber\\
    &\quad+\beta ^2\mathcal{R} \left[\Psi_{,\theta}  (\lambda\Psi-\alpha\Lambda)+\alpha\Psi  \Lambda_{,\theta}+\lambda\Lambda\Phi_{,\theta}\right]\nonumber\\
    &\quad-\lambda\beta^2\mathcal{R}_{,\theta} \left(\Lambda\Phi +\Psi ^2\right)\Big\},\nonumber\\
    \Omega_{\hat{2}}&= \pm \frac{\left(\Psi ^2+\Lambda\Phi\right)^{3/2}}{2\sqrt{\Delta  \mathcal{R} \left(\varepsilon^2 \Psi ^2+2 \varepsilon \Lambda  \ell \Psi +\Lambda ^2 \Phi +\Lambda \Psi ^2+\Lambda ^2 l^2\right)}}\nonumber\\
    &\quad\times\Big\{\Psi_{,r}\big[\varepsilon^2\left(\Psi^2-2 \Lambda\Phi\right)+4 \varepsilon\Lambda\ell\Psi +\Lambda  \left(\Lambda\Phi+\Psi^2\right)\nonumber\\
    &\quad+\Lambda ^2\ell^2\big]-\Psi\Lambda_{,r}\left[3 \varepsilon \ell \Psi-2 \varepsilon^2 \Phi  +\Lambda\Phi+\Lambda  \ell^2+\Psi ^2\right]\nonumber\\
&\quad+\varepsilon\Lambda\Phi_{,r} \left[\varepsilon \Psi+\Lambda\ell\right]
    \Big\},\nonumber\\
    \Omega_{\hat{3}}&= \pm \frac{\beta}{2 \gamma  \sqrt{\Theta\Xi}}\left[\alpha\left(\Lambda  \mathcal{R}_{,\theta}-\mathcal{R} \Lambda_{,\theta}\right)+\lambda\right(\Psi  \mathcal{R}_{,\theta}-\mathcal{R}\Psi_{,\theta} \left)\right].
\end{align}

It is worth noting here that, unlike the Schwarzschild case for which the angular velocity comes out lying along the polar direction $\hat2$, the general metric (\ref{GeneralMetric}) allows for an angular velocity with all three components nonvanishing even when the spacetime is not rotating. The reason behind this is the fact that even when the particle's motion is taking place along the equatorial plane, the time and the radial components of the metric along that plane possess vertical gradients along the polar axis $\hat2$. 

On the other hand, if the gravitational source is, in addition, also spinning along a direction that is perpendicular to its equatorial plane along which a particle happens to be moving, the frame-dragging effect and the geodetic effect would add up to give an amplified spin precession along the direction $\hat2$. If, instead, the particle's motion was taken to be along any other plane than the one that is perpendicular to the source's rotation, the two effects would be distinguishable from one another as their respective  induced angular velocities of precession would be pointing along two different directions. This is another reason behind our choice of the plane of the particle's motion since spin precession is maximal in this case. For completeness, however, we shall first consider a general planar motion of the spinning particle without restricting the latter to the equatorial plane. Showing that spin precession becomes maximum along the equatorial plane will be done in Sec.\,\ref{Sec:HelicityFlip} for each one of the spacetimes considered using the angular velocity components we obtain here for the general case.  

In what follows, we shall use these three general expressions (\ref{GeneralAngularVelocity}) to extract the components of the angular velocity three-vector of precession when the particle freely propagates along a plane perpendicular to the axis of symmetry within thirteen different spacetimes.
\subsection{The de Sitter spacetime}\label{Sec:dS}
We start our series of metrics with the de Sitter spacetime \cite{DeSitter} of positive cosmological constant $C$, written in the static coordinates $(t,r,\theta,\phi)$. In fact, while such a spacetime describes an empty spherically symmetric universe undergoing an accelerated expansion, when written in such coordinates it takes the form of a static metric that we can use to extract the metric components corresponding to the general form (\ref{GeneralMetric}). We shall therefore work with the following metric components of this spacetime\cite{deSitter2,GriffithsPodolski}:
\begin{align}\label{dSitterMetric}
    \Lambda&=1-\frac{Cr^2}{3},\! \qquad \mathcal{R}=\left(1-\frac{Cr^2}{3}\right)^{-1}\!\!\!,\qquad\! \Theta=r^2,\nonumber\\
    \Phi&=r^2\sin^2\theta\, \qquad \Psi=0.
\end{align}

This spacetime possesses a cosmological horizon that lies at $r=\sqrt{3/C}$. A spinning particle freely propagating within this spacetime would not have any orbital angular momentum per unit mass $\ell$ since there is no central gravitational source around which it would be deflected. The particle would keep coasting with the constant energy per unit mass $\varepsilon$. On the other hand, plugging the metric components (\ref{dSitterMetric}) into the general expressions (\ref{GeneralAngularVelocity}), we find that the particle's spin would precess with an angular velocity vector $\boldsymbol{\Omega}$ that has the following components:
\begin{align}\label{AngularVelocitydeSitter}
    \Omega_{\hat{1}}&=\mp\frac{ \ell  \cos \theta  \sqrt{\ell ^2+r^2 \sin ^2\theta}}{r^3\sin ^3\theta} ,\nonumber\\
    \Omega_{\hat{2}}&= \pm\frac{\varepsilon  \ell  \sin \theta}{\ell ^2+r^2 \sin ^2\theta},\nonumber\\
    \Omega_{\hat{3}}&=0.
    \end{align}
This result clearly shows that the spinning particle would not undergo any spin precession within de Sitter spacetime since setting $\ell=0$ in the first two expressions yields an angular velocity that is identically zero. 
\subsection{The Bertotti–Robinson spacetime}\label{Sec:BR}
Our second metric is that of the Bertotti–Robinson spacetime \cite{Berttoti,Robinson}, a conformally flat solution to the Einstein-Maxwell equations. This spacetime is conformally flat as it takes the following metric components in the spherical coordinates $(r,\theta,\phi)$ \cite{GriffithsPodolski}:
\begin{align}\label{BRMetric}
    \Lambda&=\frac{Q^2}{r^2}, \qquad \mathcal{R}=\frac{Q^2}{r^2}, \qquad \Theta=Q^2, \qquad \Phi=Q^2\sin^2\theta,\nonumber\\
    \Psi&=0,
\end{align}
where $Q$ is a static electric charge. The importance of this spacetime in the literature lies in its use when discussing specific regions of more well-known spacetimes, such as the region near the degenerate horizon of an extreme limit of the Reissner-Nordstr\"om spacetime \cite{Carter}. Also, despite being amenable to be written in the form of an anti-de Sitter spacetime of radius $Q$ (i.e., of negative cosmological constant $C=-3/Q^2$) using a specific change of coordinates (given explicitly in Ref.\,\cite{GriffithsPodolski}), we shall consider this spacetime in its conformally flat form (\ref{BRMetric}).

Therefore, as in the case of the de Sitter spacetime (\ref{dSitterMetric}), a spinning particle freely propagating within this spacetime would not have any orbital angular momentum per unit mass $\ell$ since there is no central gravitational source around which it would be deflected. The particle would keep coasting with the constant energy per unit mass $\varepsilon$. On the other hand, plugging the metric components (\ref{BRMetric}) into the general expressions (\ref{GeneralAngularVelocity}), we find that the particle's spin would precess with an angular velocity vector $\boldsymbol{\Omega}$ having the components:
\begin{align}\label{AngularVelocityBR}
    \Omega_{\hat{1}}&= \mp\frac{\ell \cos \theta\sqrt{\ell^2+Q^2\sin^2\theta}}{|Q|^3 \sin^3\theta},\nonumber\\
    \Omega_{\hat{2}}&=0 ,\nonumber\\
    \Omega_{\hat{3}}&=0.
    \end{align}
    Similarly to the case of de Sitter spacetime, this result shows that the spinning particle would not undergo any spin precession within the Bertotti–Robinson spacetime since setting $\ell=0$ in the first expression yields an angular velocity vector that is identically zero. 

\subsection{The Melvin-Bonnor spacetime}\label{Sec:Melvin-Bonnor}
The Melvin-Bonnor spacetime \cite{Bonnor,Melvin} is an electrovacuum static solution of Einstein's field equations with a cylindrical symmetry arising from a constant and uniform electric or magnetic field. The uniform electric or magnetic field is conventionally taken to be aligned along the $z$-direction and having a constant magnitude determined by a single parameter, customarily denoted by $B$. We use here the following metric components for such a spacetime (usually written in the literature in cylindrical coordinates \cite{GriffithsPodolski}) after expressing them in the spherical coordinates $(r,\theta,\phi)$:
\begin{align}\label{MelvinBMetric}
    \Lambda&=\mathcal{R}=\left(1+\frac{B^2r^2}{4}\sin^2\theta\right)^2\!\!\!,\quad\! \Theta=r^2\left(1+\frac{B^2r^2}{4}\sin^2\theta\right)^{2},\nonumber\\ \Phi&=r^2\sin^2\theta\left(1+\frac{B^2r^2}{4}\sin^2\theta\right)^{-2},\qquad\Psi=0.
\end{align}
When only a uniform and constant magnetic field is present, the spacetime is usually referred to as the Melvin magnetic universe. The spacetime is mainly used for modeling astrophysical processes and gravitational collapse by taking it as a remarkably stable limit of more complicated spacetimes. See Ref.\,\cite{GriffithsPodolski} for more discussions on the mathematical properties of this spacetime.

While there is no event horizon in this spacetime, and the latter is not asymptotically flat, both timelike and lightlike geodesics of particles propagating in this spacetime are bound to finite radii that depend on the parameter $B$ \cite{Thorne,Melvin-Wallingford}. We shall come back to this constraint on timelike geodesics in Sec.\,\ref{SubSec:ProbabilityInMelvinB} where we compute the spin-flip probability of neutrinos propagating within it. Here we simply compute the angular velocity of spin precession of freely propagating neutral spinning particles.

A spinning particle freely propagating within this spacetime would have a conserved orbital angular momentum per unit mass $\ell$ since there is the possibility of the particle being deflected around the central axis of symmetry. The particle would then be deflected while conserving its constant energy per unit mass $\varepsilon$. Plugging the metric components (\ref{MelvinBMetric}) into the general expressions (\ref{GeneralAngularVelocity}), we find that the particle's spin would precess with an angular velocity vector $\boldsymbol{\Omega}$ that has the following components:
\begin{align}\label{AngularVelocityMelvinB}
\Omega_{\hat{1}}&=\mp\frac{\left(4-3 B^2 r^2 \sin ^2\theta\right)\ell\cos\theta}{4\left(4+B^2 r^2 \sin ^2\theta\right)r^3\sin\theta}\nonumber\\
&\qquad\times\left[B^4 \ell ^2 r^4+\frac{8r^2\left(B^2 \ell ^2+2\right)}{\sin^2\theta}+\frac{16 \ell ^2}{\sin^4\theta}\right]^{\frac{1}{2}},\nonumber\\
\Omega_{\hat{2}}&=\pm\frac{64 \varepsilon \ell  \left(4-B^2 r^2 \sin ^2\theta\right)}{\left(4+B^2 r^2 \sin ^2\theta\right)^2\sin^3\theta}\nonumber\\
&\qquad\times\left[B^4 \ell ^2 r^4+\frac{8r^2\left(B^2 \ell ^2+2\right)}{\sin^2\theta}+\frac{16 \ell ^2}{\sin^4\theta}\right]^{-1},\nonumber\\
    \Omega_{\hat{3}}&=0.
    \end{align}
This result shows that a spinning particle that does not get deflected in this spacetime would not undergo any spin precession either since setting $\ell=0$ in the first two expressions also yields an angular velocity that is identically zero. 

\subsection{The Reissner-Nordstr\"om spacetime}\label{Sec:RN}
The other static electrovacuum solution of Einstein's field equations that we consider in this paper is the Reissner-Nordstr\"om spacetime \cite{Reissner,Weyl,Nordstrom}. This spacetime belongs to the Weyl class of electrovacuum solutions \cite{GriffithsPodolski}. This solution describes the spacetime around a charged black hole of mass $M$ and of static electric charge $Q$. We use the following metric components in spherical coordinates $(r,\theta,\phi)$  \cite{GriffithsPodolski}:
\begin{align}\label{RNMetric}
    \Lambda&=\left(1-\frac{r_s}{r}+\frac{Q^2}{r^2}\right),\! \quad \mathcal{R}=\left(1-\frac{r_s}{r}+\frac{Q^2}{r^2}\right)^{-1}\!\!\!,\quad\! \Theta=r^2,\nonumber\\
    \Phi&=r^2\sin^2\theta,\quad \Psi=0.
\end{align}
This spacetime is asymptotically flat and it has two event horizons which are located at $r_{\pm}=M\pm\sqrt{M^2-Q^2}$. We assume that our particle is moving outside the outer event horizon, such that $r>r_{+}$, and that the particle gets deflected when it is close to the gravitational source.

Our spinning particle will be taken to be freely propagating within this spacetime with a conserved orbital angular momentum per unit mass $\ell$ and a conserved energy per unit mass $\varepsilon$. Plugging the metric components (\ref{RNMetric}) into the general expressions (\ref{GeneralAngularVelocity}), we find that the particle's spin would precess with an angular velocity vector $\boldsymbol{\Omega}$ having the components:
\begin{align}\label{AngularVelocityRN}
    \Omega_{\hat{1}}&=\mp\frac{ \ell\cos\theta\sqrt{\ell ^2+r^2\sin^2\theta}}{r^3\sin^3\theta},\nonumber\\
\Omega_{\hat{2}}&=\pm\frac{\varepsilon \ell\sin\theta}{\ell^2+r^2\sin^2\theta},\nonumber\\
    \Omega_{\hat{3}}&=0.
    \end{align}
For spin precession to occur, this angular velocity requires a nonvanishing orbital angular momentum per unit mass $\ell$ which is, indeed, guaranteed by the central charged gravitational source.
\subsection{The interior Schwarzschild spacetime}\label{Sec:InSchw}
To take into account also the possibility of spinning particles deflected by massive gravitational sources that are not black holes, and therefore allow for the possibility of having those particles propagate even inside the matter making those massive bodies, we consider here the interior Schwarz-\\
schild solution \cite{InSchwar}. It is an idealized spherically symmetric spacetime corresponding to the inside of a uniformly distributed mass $M$ in a spherical region of radius $R$.
We use the following metric components in the spherical coordinates $(r,\theta,\phi)$ \cite{GriffithsPodolski}:
\begin{align}\label{InSch}
    \Lambda&=\left[\frac{3}{2}\left(1-\frac{r_s}{R}\right)^{\frac{1}{2}}-\frac{1}{2}\left(1-\frac{r_sr^2}{R^3}\right)^{\frac{1}{2}}\right]^2\!\!,\nonumber\\ 
    \mathcal{R}&=\left(1-\frac{r_sr^2}{R^3}\right)^{-1}\!, \quad
    \Theta=r^2, \quad \Phi=r^2\sin^2\theta, \quad
    \Psi=0,
\end{align}
where $r_s=2M$. This spacetime has no singularity and no event horizon, and it is everywhere regular inside the spherical region of radius $R$.

Even though the spinning particle is now taken to be moving inside the massive body, we shall assume that the particle is not experiencing any other interaction with the matter making the body other than the gravitational interaction.  Our spinning particle will then also be taken to be freely propagating within this spacetime with a conserved energy per unit mass $\varepsilon$ and a conserved orbital angular momentum per unit mass $\ell$ as it gets deflected inside the body.

Plugging the metric components (\ref{InSch}) into the general expressions (\ref{GeneralAngularVelocity}), we find that the particle's spin would precess with the following components of the angular velocity vector $\boldsymbol{\Omega}$:
\begin{align}\label{AngularVelocityInSch}
    \Omega_{\hat{1}}&=\mp\frac{ \ell  \cos \theta \sqrt{\ell ^2+r^2 \sin ^2\theta}}{r^3\sin^3\theta} ,\nonumber\\
\Omega_{\hat{2}}&=\mp \frac{2 \varepsilon\ell\sin \theta}{\ell ^2+r^2 \sin ^2\theta}\left(1-3R \sqrt{\frac{R-r_s}{R^3-r_s r^2}}\right)^{-1}  ,\nonumber\\
    \Omega_{\hat{3}}&=0.
    \end{align}
Spin precession is here also allowed since the particle necessarily gets deflected inside the body giving the particle a nonvanishing angular momentum per unit mass $\ell$.
\subsection{The Hayward spacetime}\label{Sec:Hay}
The other spacetime that is static, spherically symmetric, singularity-free and everywhere regular that we consider is the Hayward spacetime \cite{Hayward}. It is made of a central mass $M$ that is embedded in an expanding universe of positive cosmological constant $C$. We use the following metric components in the spherical coordinates $(r,\theta,\phi)$ \cite{Hayward}:
\begin{align}\label{HaywardMetric}
    \Lambda&=1-\frac{r_sr^2}{r^3+\frac{1}{3}Cr_s}, \quad \mathcal{R}=\left(1-\frac{r_sr^2}{r^3+\frac{1}{3}Cr_s}\right)^{-1}\!\!\!,\quad
    \Theta=r^2,\nonumber\\
    \Phi&=r^2\sin^2\theta, \quad
    \Psi=0,
\end{align}
where $r_s=2M$. This spacetime has no event horizon when $M<\frac{9}{4}C$, it has a single horizon when $M=\frac{9}{4}C$ and it has two distinct horizons when $M>\frac{9}{4}C$. The two horizons lie at $r_-\approx r_s$ and $r_+\approx \sqrt{3/C}$, respectively, for $M\gg\frac{9}{4}C$ and they coincide with each other for $M=\frac{9}{4}C$ \cite{Hayward}.

Our spinning particle is also allowed in this case to be deflected in this spacetime as it freely propagates around the central mass with a conserved orbital angular momentum per unit mass $\ell$ and a conserved energy per unit mass $\varepsilon$. Plugging the metric components (\ref{HaywardMetric}) into the general expressions (\ref{GeneralAngularVelocity}), we find that the particle's spin would precess with an angular velocity vector $\boldsymbol{\Omega}$ that has the components:
\begin{align}\label{AngularVelocityHayward}
    \Omega_{\hat{1}}&=\mp\frac{ \ell  \cos \theta \sqrt{\ell ^2+r^2 \sin ^2\theta}}{r^3\sin^3\theta},\nonumber\\
    \Omega_{\hat{2}}&=\pm\frac{\varepsilon \ell  \sin \theta}{\ell ^2+r^2 \sin ^2\theta} ,\nonumber\\
    \Omega_{\hat{3}}&=0.
    \end{align}
The particle necessarily gets deflected around the central mass if it is allowed to get close enough to the latter, acquiring therefore a nonvanishing angular momentum per unit mass $\ell$. Spin precession of the particle is then also allowed in this spacetime.
\subsection{The Bardeen spacetime}\label{Sec:Bardeen}
The third black hole solution that is singularity-free that we consider in this paper is the Bardeen spacetime \cite{Bardeen}. This solution describes the spacetime around a point mass $M$ carrying a magnetic charge $q$. We use the following metric components in the spherical coordinates $(r,\theta,\phi)$:
\begin{align}\label{BardeenMetric}
    \Lambda&=1-\frac{r_sr^2}{(q^2+r^2)^{\frac{3}{2}}}, \quad \mathcal{R}=\left[1-\frac{r_sr^2}{(q^2+r^2)^{\frac{3}{2}}}\right]^{-1}\!\!\!,\quad \Theta=r^2,\nonumber\\
    \Phi&=r^2\sin^2\theta,\quad
    \Psi=0,
\end{align}
where $r_s=2M$. This spacetime is regular at $r=0$. It also has an event horizon at $r=r_+$ which is the positive root of the equation $\Lambda=0$. This spacetime is not asymptotically flat though.

We let our spinning particle freely propagate within this spacetime and be deflected with a conserved orbital angular momentum per unit mass $\ell$ and a conserved energy per unit mass $\varepsilon$. Plugging the metric components (\ref{BardeenMetric}) into the general expressions (\ref{GeneralAngularVelocity}), we find that the particle's spin would precess with an angular velocity vector $\boldsymbol{\Omega}$ that has the following components:

\begin{align}\label{AngularVelocityBardeen}
    \Omega_{\hat{1}}&=\mp\frac{ \ell  \cos \theta \sqrt{\ell ^2+r^2 \sin ^2\theta}}{r^3\sin^3\theta},\nonumber\\
    \Omega_{\hat{2}}&=\pm\frac{\varepsilon \ell  \sin \theta}{\ell ^2+r^2 \sin ^2\theta} ,\nonumber\\
    \Omega_{\hat{3}}&=0.
    \end{align}
A spinning particle that acquires an angular momentum per unit mass $\ell$ after being deflected around the central mass will therefore experience a spin precession in this spacetime.
\subsection{The Kiselev spacetime}\label{Sec:Kiselev}
One of the black hole solutions we consider in this paper is taken from the class of Kiselev’s black holes embedded in a quintessential spacetime \cite{Kiselev}. We use the following metric components in the spherical coordinates $(r,\theta,\phi)$, describing a black hole of mass $M$ embedded in a universe undergoing a quintessence-driven expansion:
\begin{align}\label{KiselevMetric}
    \Lambda&=1-\frac{r_s}{r}-\left(\frac{r_q}{r}\right)^{3\omega_q+1}, \quad \mathcal{R}=\left[1-\frac{r_s}{r}-\left(\frac{r_q}{r}\right)^{3\omega_q+1}\right]^{-1}\!\!\!,\nonumber\\
    \Theta&=r^2, \quad \Phi=r^2\sin^2\theta, \quad \Psi=0,
\end{align}
where $r_s=2M$, $\omega_q$ is the quintessential state parameter such that $-1<\omega_q<-\frac{1}{3}$, and $r_q$ is a normalization constant. For the extreme case $\omega_q=-1$, one recovers the Schwarzschild spacetime when setting $r_q=0$ in Eq.\,(\ref{KiselevMetric}). One recovers from this spacetime the Schwarzschild-de Sitter spacetime of positive cosmological constant $C$ when setting $r_q=\sqrt{3/C}$. The spacetime solution (\ref{KiselevMetric}) is therefore more flexible than the Schwarzschild-de Sitter spacetime as it allows also for studying black holes within quintessential models of cosmology. 

We let our spinning particle freely propagate within this spacetime and be deflected by the black hole with a conserved orbital angular momentum per unit mass $\ell$ and a conserved energy per unit mass $\varepsilon$. Plugging the metric components (\ref{KiselevMetric}) into the general expressions (\ref{GeneralAngularVelocity}), we find that the particle's spin would precess with an angular velocity vector $\boldsymbol{\Omega}$ having the components:

\begin{align}\label{AngularVelocityKiselev}
    \Omega_{\hat{1}}&=\mp\frac{ \ell  \cos \theta \sqrt{\ell ^2+r^2 \sin ^2\theta}}{r^3\sin^3\theta},\nonumber\\
    \Omega_{\hat{2}}&=\pm\frac{\varepsilon \ell  \sin \theta}{\ell ^2+r^2 \sin ^2\theta} ,\nonumber\\
    \Omega_{\hat{3}}&=0.
    \end{align}
These components' expressions are identical to those obtained above for the de Sitter, the Reissner-Nordstr\"om, the Hayward and the Bardeen spacetimes. The reason is that these metrics have all in common that their components are of the form $\Lambda=\Lambda(r)$, $\mathcal{R}=\mathcal{R}(r)$, $\Theta=r^2$, $\Phi=r^2\sin^2\theta$ and $\Psi=0$. Therefore, a spinning particle that freely propagates in all these four spacetimes and gets deflected acquires an angular momentum per unit mass $\ell$ that will make it experience spin precession with an angular velocity given by the unique result (\ref{AngularVelocityKiselev}).
\subsection{The Schwarzschild-Melvin spacetime}\label{Sec:SchwarMel}
The Schwarzschild-Melvin metric describes the spacetime around a neutral point mass $M$ that is immersed inside a constant and uniform electric or magnetic field of strength $B$. In the absence of the point mass $M$, this constant electric or magnetic field is what gives rise to the Melvin-Bonnor spacetime (\ref{MelvinBMetric}) we considered above. The Schwarzschild-Melvin solution is the electrovacuum spacetime solution to the Einstein field equations that belongs to the more general Ernst solution \cite{Ernst} that describes an accelerating and charged black hole immersed inside a constant and uniform electric or magnetic field of strength $B$. We work here with the following metric components in the $(r,\theta,\phi)$ coordinates \cite{GriffithsPodolski}:
\begin{align}\label{SchwarMel}
\Lambda&=\left(1+\frac{B^2r^2}{4}\sin^2\theta\right)^2\left(1-\frac{r_s}{r}\right),\nonumber\\
\mathcal{R}&=\left(1+\frac{B^2r^2}{4}\sin^2\theta\right)^2\left(1-\frac{r_s}{r}\right)^{-1},\nonumber\\
    \Theta&=r^2\left(1+\frac{B^2r^2}{4}\sin^2\theta\right)^2,\nonumber\\
\Phi&=r^2\sin^2\theta\left(1+\frac{B^2r^2}{4}\sin^2\theta\right)^{-2}, \qquad
    \Psi=0,
\end{align}
where $r_s=2M$. This spacetime has a single event horizon at $r=r_s$ and the spacetime is not asymptotically flat. Even though the particles' trajectories are not integrable (i.e., are chaotic) \cite{Karas}, we shall assume here, for the sake of simplicity and definiteness, a well-defined geodesic trajectory for our neutral spinning particle.

We thus assume that the neutral spinning particle is deflected in this spacetime as it freely propagates around the central mass $M$ with a conserved orbital angular momentum per unit mass $\ell$ and a conserved energy per unit mass $\varepsilon$. Plugging the metric components (\ref{SchwarMel}) into the general expressions (\ref{GeneralAngularVelocity}), we find that the particle's spin would precess with an angular velocity vector $\boldsymbol{\Omega}$ that has the following components:
\begin{align}\label{AngularVelocitySchwarMel}
    \Omega_{\hat{1}}&=\mp\frac{\ell\cos\theta  \left(4-3B^2r^2 \sin^2 \theta\right)}{4\left(4+B^2 r^2\sin^2\theta\right)r^3\sin\theta}\nonumber\\
    &\quad\times\left[B^4 \ell ^2 r^4+\frac{8r^2\left(B^2 \ell ^2+2\right)}{\sin^2\theta}+\frac{16 \ell ^2}{\sin^4\theta}\right]^{\frac{1}{2}}, \nonumber\\
    \Omega_{\hat{2}}&=\pm\frac{64 \varepsilon \ell  \left(4-B^2r^2 \sin^2 \theta\right)}{\left(4+B^2 r^2\sin^2\theta\right)^2\sin^3\theta}\nonumber\\
    &\quad\times\left[B^4 \ell ^2 r^4+\frac{8r^2\left(B^2 \ell ^2+2\right)}{\sin^2\theta}+\frac{16 \ell ^2}{\sin^4\theta}\right]^{-1} ,\nonumber\\
    \Omega_{\hat{3}}&=0.
    \end{align}
This velocity three-vector does not vanish only for particles of nonzero angular momentum per unit mass $\ell$. 
\subsection{The Weyl spacetime}\label{Sec:Weyl}
Among the vast class of Weyl spacetimes \cite{Weyl}, we consider here the one that arises from a neutral central point mass $M$. Unlike the Schwarzschild spacetime, though, this spacetime is not spherically symmetric. Like the Schwarzschild spacetime, it has a singularity at $r=0$, but unlike the Schwarzschild spacetime it has no event horizon. From the simplest member of the Weyl class in the spherical coordinates $(r,\theta,\phi)$, we pick up the following spacetime of metric components \cite{GriffithsPodolski}:
\begin{align}\label{WeylMetric}
    \Lambda&=\exp\left(-\frac{r_s}{r}\right), \qquad \mathcal{R}=\exp\left(\frac{r_s}{r}-\frac{r_s^2}{4r^2}\sin^2\theta\right),\nonumber\\
\Theta&=r^2\exp\left(\frac{r_s}{r}-\frac{r_s^2}{4r^2}\sin^2\theta\right),\quad \Phi=r^2\sin^2\theta\exp\left(\frac{r_s}{r}\right),\nonumber\\
\Psi&=0,
\end{align}
where $r_s=2M$. The absence of a horizon in this spacetime makes the curvature singularity at $r=0$ a naked singularity. Nevertheless, as any radiation that would be emitted from it would be infinitely red-shifted, the naked singularity is said to be ``invisible'' (see Ref.\,\cite{GriffithsPodolski} and references therein).

We let our spinning particle freely propagate within this spacetime and be deflected thanks to the central mass with a conserved orbital angular momentum per unit mass $\ell$ and a conserved energy per unit mass $\varepsilon$. Plugging the metric components (\ref{WeylMetric}) into the general expressions (\ref{GeneralAngularVelocity}), we find that the particle's spin would precess with an angular velocity vector $\boldsymbol{\Omega}$ having the components:
\begin{align}\label{AngularVelocityWeylMetric}
    \Omega_{\hat{1}}&=\mp\frac{\ell  \sin 2\theta  \exp\left(\frac{r_s^2 \sin^2\theta
    }{8 r^2}-\frac{3 r_s
    }{2 r}\right)}{8 r^5 \sqrt{\ell ^2+r^2 \sin ^2\theta \exp \left(\frac{r_s}{r}\right)}}\nonumber\\
    &\quad\times\Bigg{\{}\left[1\!-\!\varepsilon^2 \exp \left(\frac{r_s}{r}\right)\right]r_s^2 r^2 \exp \left(\frac{r_s}{r}\right)\nonumber\\
    &\qquad\qquad\qquad\qquad+\frac{4 r^4 \exp \left(\frac{r_s}{r}\right)+r_s^2 \ell ^2}{\sin ^2\theta}+\frac{4 r^2 \ell ^2}{\sin^4 \theta}\Bigg{\}} ,\nonumber\\
\Omega_{\hat{2}}&=\mp\frac{\varepsilon (r_s-2 r) \ell\sin \theta }{2 r \left[\ell ^2+r^2\sin ^2\theta\exp \left(\frac{r_s}{r}\right)\right]}\exp \left(\frac{r_s^2 \sin ^2\theta
    }{8 r^2}+\frac{r_s
    }{2 r}\right) ,\nonumber\\
\Omega_{\hat{3}}&=\mp \frac{\varepsilon  r_s^2 \sin 2 \theta}{8 r^3} \exp \left(\frac{r_s^2 \sin ^2\theta
    }{8 r^2}+\frac{r_s
    }{2 r}\right)\nonumber\\
    &\quad\times\left[\frac{\varepsilon^2 r^2 \sin ^2\theta \exp \left(\frac{2 r_s}{r}\right)-r^2 \sin ^2\theta \exp \left(\frac{r_s}{r}\right)-\ell ^2}{r^2 \sin ^2\theta \exp \left(\frac{2r_s}{r}\right)+\ell ^2\exp \left(\frac{r_s}{r}\right)}\right]^\frac{1}{2}.
\end{align}
We immediately notice from these expressions that even for $\ell=0$, the angular velocity does not vanish identically. This result therefore shows that, in contrast to all the spacetimes we have dealt with so far, the Weyl spacetime would induce a spin precession even on particles that move undeflected within it.

\subsection{The Kerr spacetime}\label{Sec:InKerr}
For the case of a spinning gravitational source, we consider the Kerr spacetime \cite{Kerr}. This black hole solution describes spacetime around a spinning spherical massive body of mass $M$ and of angular momentum $J$. We use the following metric components in the spherical coordinates $(r,\theta,\phi)$ \cite{GriffithsPodolski}:
\begin{align}\label{KerrMetric}
    \Lambda&=\frac{A-a^2\sin^2\theta}{\rho^2}, \qquad \mathcal{R}=\frac{\rho^2}{A}, \qquad \Theta=\rho^2,\nonumber\\
    \Phi&=\frac{(r^2+a^2)^2-Aa^2}{\rho^2}\sin^2\theta,\nonumber\\
    \Psi&=\frac{(r^2+a^2)a-aA}{\rho^2}\sin^2\theta,
\end{align}
where $\rho^2=r^2+a^2\cos^2\theta$, $A=r^2-r_sr+a^2$, $r_s=2M$ and $a=J/M$. This metric is not only suitable for studying spinning black holes, but also for studying most of the astrophysical compact objects of interest, including fast rotating neutron stars \cite{Cipolletta}.

Given that when plugging the metric components (\ref{KerrMetric}) into the general expression (\ref{GeneralAngularVelocity}), the resulting components of the angular velocity of spin precession take on extremely long expressions, we display the three components of the vector $\boldsymbol{\Omega}$ for the general case in Appendix \ref{Sec:AppB}. Here, we display only the components of the angular velocity of spin precession when the particle is moving along the equatorial plane, i.e., for $\theta=\pi/2$:
\begin{align}\label{Pi/2AngularVelocityKerr}
&\Omega_{\hat{1}}=0,\nonumber\\
&\Omega_{\hat{2}}=\mp\Bigg{\{}a^3 r_s \left[r-\varepsilon ^2 (3 r+\mathcal{X})\right]+a^2 \varepsilon  \ell  r_s (3 r+2 \mathcal{X})\nonumber\\
&\quad\qquad-a r_s \left[r^2 \left(\mathcal{X}-2 \varepsilon ^2 [\mathcal{X}-r]\right)+\mathcal{X} \ell ^2\right]-2 \varepsilon  r^2 \mathcal{X}^2 \ell \Bigg{\}}\nonumber\\
&\qquad\times\Bigg{\{}{2 r^2 \left[a^2 \left(\varepsilon ^2 r_s^2-r \mathcal{X}\right)-2 a \varepsilon  \mathcal{X} \ell  r_s+\mathcal{X}^2 \left(r^2+\ell ^2\right)\right]}\Bigg{\}}^{-1},\nonumber\\
    &\Omega_{\hat{3}}=0.
\end{align}
Here, we set $\mathcal{X}=r_s-r$.

The fact that even for $\ell=0$ the angular velocity of spin precession does not vanish identically within this spacetime is easy to understand. It is the result of the well-known frame-dragging effect of spacetime. This spacetime thus induces spin precession even on particles for which their initial angular momentum per unit mass $\ell$ is zero. The precession is induced by the spacetime by merely making the particle follow the ``flow'' of the latter.

\subsection{The wormhole spacetime}\label{Sec:Wormhole}
Wormholes, even though they have never been observed in Nature, are very ubiquitous in the spacetime physics literature. Many variants of different complexities have thus been introduced in the literature. For the sake of simplicity, however, but without losing the generic features of wormholes and the possibility of having a regular one, we consider here the static wormhole solution introduced in Ref.\,\cite{Teo}. Its metric components written in the spherical coordinates $(r,\theta,\phi)$ read:
\begin{align}\label{SWHMetric}
    \Lambda&=1-\frac{r_s}{\sqrt{r^2+\alpha^2}}, \qquad \mathcal{R}=\left(1-\frac{r_s}{\sqrt{r^2+\alpha^2}}\right)^{-1},\nonumber\\ \Theta&=r^2+\alpha^2,\!\!\!\qquad \Phi=\left(r^2+\alpha^2\right)\sin^2\theta,\!\!\!\qquad \Psi=0,
\end{align}
where $r_s=2M$ and $\alpha$ is a free parameter that is used to regularize the central singularity of the wormhole. 
Plugging the metric components (\ref{SWHMetric}) into the general expressions (\ref{GeneralAngularVelocity}), we find that the particle's spin would precess with an angular velocity vector $\boldsymbol{\Omega}$ of components:
\begin{align}\label{AngularVelocitySWH}
    \Omega_{\hat{1}}&=\mp\frac{ \ell  \cos \theta}{\left(r^2+\alpha ^2\right)^{3/2} \sin ^2\theta}\sqrt{r^2+\alpha ^2+\frac{\ell ^2}{\sin ^2\theta}},\nonumber\\
    \Omega_{\hat{2}}&= \pm\frac{\varepsilon r \ell  \sin \theta}{\sqrt{\alpha ^2+r^2} \left[\sin ^2\theta \left(\alpha ^2+r^2\right)+\ell ^2\right]},\nonumber\\
    \Omega_{\hat{3}}&=0 .
    \end{align}
The appearance of the angular momentum per unit mass $\ell$ of the particle also as a multiplicative factor in these components shows that an undeflected freely propagating  spinning particle in this spacetime would experience no spin precession.
\subsection{The straight spinning cosmic string spacetime
}\label{Sec:InStraightSpinningString
}

The last spacetime metric we consider in this paper is that of a straight spinning cosmic string. Although simple, the spacetime metric we shall work with here captures the main interesting features of spacetime around a straight cosmic string when the latter is spinning \cite{Mazur}. We use the following metric components written in the spherical coordinates $(r,\theta,\phi)$:
\begin{align}\label{SSCSMetric}
    \Lambda&=1,  \quad \mathcal{R}=1,\quad \Theta=r^2,\quad \Phi=a^2r^2\sin^2\theta-16J^2,\nonumber\\
    \Psi&=4J,
\end{align}
where $a$ is a strictly positive real parameter related to the mass per unit length $\mu$ of the string by $a=1-4\mu$, and $J$ is the angular momentum per unit length of the spinning string. 
Plugging the metric components (\ref{SSCSMetric}) into the general expressions (\ref{GeneralAngularVelocity}), we find that the particle's spin would precess with an angular velocity vector $\boldsymbol{\Omega}$ that has the components:
\begin{align}\label{AngularVelocitySSCS}
    \Omega _{\hat{1}}&=\mp\frac{ (4 \varepsilon J+\ell )\cos\theta \sqrt{(4 \varepsilon J+\ell )^2+a^2 r^2\sin^2\theta}}{a^2 r^3\sin^3 \theta},\nonumber\\
    \Omega _{\hat{2}}&=\pm\frac{a \varepsilon (4 \varepsilon J+\ell) \sin \theta}{(4 \varepsilon J+\ell)^2+a^2 r^2 \sin ^2\theta},\nonumber\\
    \Omega _{\hat{3}}&=0.
\end{align}
It is clear from these expressions that, in contrast to the spinning wormhole above, even an undeflected freely propagating spinning particle would experience a spin precession within this spacetime. 
\section{Neutrino helicity flip in various spacetimes}\label{Sec:HelicityFlip}
Now that we have the angular velocities of the spin precession underwent by a spinning particle moving within various spacetimes, we can apply our results to neutrinos to find the probability that the latter flip their helicity as they freely propagate within those spacetimes. The probability for neutrinos helicity flip can be obtained using the effective Hamiltonian $H_{\rm eff}(\textbf{r})$ that arises from the spin-gravity interaction. The effective Hamiltonian $H_{\rm eff}(\textbf{r})$ is, in turn, obtained from the three Pauli matrices $\boldsymbol\sigma=(\sigma_1,\sigma_2,\sigma_3)$ and the spin precession angular velocity $\boldsymbol{\Omega}$ as, $H_{\rm eff}(\textbf{r})=\frac{1}{2}{\boldsymbol\sigma}.{\boldsymbol{\Omega}}$. For an initial spin state $\ket{S_{\rm in}}$, the probability for a neutrino to be found in the final spin state $\ket{S_{\rm fi}}$ is given by \begin{equation}\label{GeneralProbability}
\mathcal{P}(S_{\rm in}\rightarrow S_{\rm fi})=\left|\bra{S_{\rm fi}}\textbf{T}\exp\left[-i\int_{\tau_i}^{\tau_f}H_{\rm eff}(\textbf{r}){\rm d}\tau\right]\ket{S_{\rm in}}\right|^2,
\end{equation}
where $\textbf{T}$ is the usual time-ordering operator and $\tau_i$ and $\tau_f$ are the initial and final proper times. For an initially left-handed neutrino moving with a general four-velocity $u^\mu$ in spherical coordinates in the equatorial plane, the normalized initial spin state can be written as  $\ket{S_{\rm in}}=
     \frac{1}{\sqrt{2}}(-1,\xi e^{i\phi})$,
where $\xi=(u^1+iru^3)/\sqrt{(u^1)^2+(ru^3)^2}$. The final right-handed state of the neutrino then reads $\ket{S_{\rm fi}}=\frac{1}{\sqrt{2}}(1,\xi e^{i\phi})$.
For these initial and final spin states, formula (\ref{GeneralProbability}) yields the following probability for a left-handed neutrino $\ket{\nu_L}$ to transform into a right-handed neutrino $\ket{\nu_R}$ \cite{NSO}:
\begin{align}\label{OrbitingSpinProbability}
&\mathcal{P}(\ket{\nu_L}\rightarrow \ket{\nu_R})=\left[\int_{\tau_i}^{\tau_f}\,\Omega(\textbf{r})\,{\rm d}\tau\right]^{-2}\sin^2\left[\tfrac{1}{2}\int_{\tau_i}^{\tau_f}\Omega(\textbf{r})\,{\rm d}\tau\right]\nonumber\\
&\times\Bigg(\left[\int_{\tau_i}^{\tau_f}\,\mathfrak{Im}\!\left(\xi\left[\Omega_{\hat 1}(\textbf{r})-i\Omega_{\hat 3}(\textbf{r})\right]\right){\rm d}\tau\right]^2\nonumber\\
&\qquad\qquad\qquad\qquad\qquad\qquad\qquad+\left[\int_{\tau_i}^{\tau_f}\,\Omega_{\hat 2}(\textbf{r})\,{\rm d}\tau\right]^2\Bigg).
\end{align}
Here, $\Omega(\textbf{r})$ is the magnitude of the precession angular velocity vector, and the imaginary term $\mathfrak{Im}(\xi[\Omega_{\hat 1}(\textbf{r})-i\Omega_{\hat 3}(\textbf{r})])$ is extracted from the complex expression $\xi[\Omega_{\hat 1}(\textbf{r})-i\Omega_{\hat 3}(\textbf{r})]$. We shall now apply this formula to extract the helicity flip probability for neutrinos propagating in the various spacetimes we considered in Sec.\,\ref{Sec:InStationaryAST}.

To be able to perform the integration in the formula (\ref{OrbitingSpinProbability}), we need to transform those integrals over the proper time $\tau$ into integrals over the radial coordinate $r$ by expressing ${\rm d}\tau$ in terms of ${\rm d}r$. To achieve that, we make use of the general expression of the four-velocity (\ref{General4Velocity}) from which it follows that ${\rm d \tau}= \pm\gamma\,{\rm d}r/\beta$. The ($\pm$) signs come from the fact that the proper time element ${\rm d}\tau$ is always positive whereas the radial coordinate element ${\rm d}r$ is positive when the particle is moving away from the gravitational source and negative when the particle is moving toward the source. In fact, the coordinate radius $r$ decreases
when the particle moves toward the gravitational source and reaches a minimum value $r_0$ at the deflection region where we have ${\rm d}r/{\rm d}\tau=0$. The coordinate $r$ starts then to increase as the particle moves away from the
gravitational source. For the case when the particle is deflected, we allow the radial coordinate $r$ to vary from $r=r_0$ to $r=r_f$, where $r_f=+\infty$ for most of our spacetime metrics from Sec.\,\ref{Sec:Precession}, except for the few ones that impose bounded orbits on timelike geodesics. When the particle is not deflected, we let the radial coordinate $r$ vary from $r=0$ to $r=+\infty$ whenever a central singularity is absent. 

Also, for undeflected particles with unbounded orbits, we need to take into account the fact that, unlike the coordinate time $t$ that would vary from $t=-\infty$ to $t=+\infty$, the proper time $\tau$ varies from $\tau=0$ to $\tau=\infty$. Therefore, we should multiply by a factor of $2$ the integrand as we switch the integral over $\tau$ to an integral over the radial coordinate $r$ for such cases. Furthermore, for bounded trajectories the initial and final radial positions $r_i$ and $r_f$ will, in all of the cases, be taken to represent large separation distances. The probability formula (\ref{OrbitingSpinProbability}) takes then the following form:
\begin{align}\label{GeneralProbabilityFormula}
&\mathcal{P}(\ket{\nu_L}\rightarrow \ket{\nu_R})\!=\!\left[\int_{r_i}^{r_f}\,\frac{\gamma\,\Omega(\textbf{r})}{\beta}\,{\rm d}r\right]^{-2}\!\!\sin^2\left[\int_{r_i}^{r_f}\frac{\gamma\,\Omega(\textbf{r})}{\beta}\,{\rm d}r\right]\nonumber\\
&\times\Bigg(\left[\int_{r_i}^{r_f}\,\mathfrak{Im}\!\left(\xi\left[\Omega_{\hat 1}(\textbf{r})-i\Omega_{\hat 3}(\textbf{r})\right]\right)\frac{\gamma}{\beta}{\rm d}r\right]^2\nonumber\\
&\qquad\qquad\qquad\qquad\qquad\qquad\qquad+\left[\int_{r_i}^{r_f}\frac{\gamma\,\Omega_{\hat 2}(\textbf{r})}{\beta}{\rm d}r\right]^2\Bigg).
\end{align}
We shall apply this formula to the various spacetimes we considered in the previous section that induce spin precession. When applying this formula to the case of deflected particles, we only need to integrate from the closest approach $r_0$ to $r_f$ and multiply by a factor of $2$ the integrand of each of the four integrals.

On the other hand, as the angular velocity components, as well as $\xi$, $\gamma$ and $\beta$, are all dependent on both $\ell$ and $\varepsilon$, we shall consider throughout this section energies per unit mass $\varepsilon$ of the neutrinos to be as high as $\varepsilon \sim 10^8$, which is a typical minimum value for supernovae neutrinos (see, for example Ref.\,\cite{NeutrinoBook2} and
the references therein), and as low as $\varepsilon\sim1$ in order to also give room to the neutrinos of the cosmic neutrino background \cite{Dey}. We then let the radial variable $r$ vary from its closest-approach value $r=r_0$ to $r=+\infty$, where for definiteness, we identify the closest approach radius $r_0$ with the impact parameter $b=\ell/\sqrt{\varepsilon^2-1}$.



\subsection{Spin-flip probability in the Melvin-Bonnor spacetime}\label{SubSec:ProbabilityInMelvinB}
As it was discussed below Eq.\,(\ref{General4Velocity}), spin precession is maximum when the particle is moving along the equatorial plane of the gravitational source. This can be readily verified for the Melvin-Bonnor metric (\ref{MelvinBMetric}). Indeed, according to the angular velocity (\ref{AngularVelocityMelvinB}) we found for spin precession of spinning particles freely propagating in this spacetime, the particles' spin precession would indeed have the largest angular speed for $\theta=\pi/2$; i.e., when the particle is moving along the equatorial plane around the axis of symmetry of the electric/magnetic field. This is readily checked by computing the derivative of the angular velocity's magnitude $\Omega$ with respect to $\theta$ using its two nonvanishing components given in Eq.\,(\ref{AngularVelocityMelvinB}). Such a derivative is indeed found to vanish for $\theta=\pi/2$. The effect of gravity on spin precession is, of course, the only contribution taken into account here.

Setting $\theta=\pi/2$ in the expressions (\ref{AngularVelocityMelvinB}) yields a single nonvanishing component for the angular velocity three-vector that reads
\begin{equation}\label{EquatorAngularVelocityMelBonnor}
\Omega_{\hat{2}}=\mp\frac{64 \varepsilon\ell  \left(B^2 r^2-4\right)}{\ell ^2 \left(B^2 r^2+4\right)^4+16 r^2 \left(B^2 r^2+4\right)^2}.
    \end{equation}
Plugging this result into the general probability expression (\ref{GeneralProbabilityFormula}), we find the following probability for a left-handed neutrino to turn into a right-handed neutrino as it propagates and gets deflected in this spacetime:
\begin{equation}\label{EquatorMelBonnorSpinProbabilityIntegral}
\mathcal{P}(\ket{\nu_L}\rightarrow \ket{\nu_R})=\sin^2\left[ \int_{x_0}^{\sqrt{\varepsilon^2-1}} \mathcal{I}(x)\,{\rm d}x \right].
\end{equation}
The function $\mathcal{I}(x)$ is obtained after performing the change of variables $x=\ell/r$ and setting $k=B\ell$. It reads
\begin{equation}\label{eq:MelvinIntegrand}
    \mathcal{I}(x)=\frac{128 \varepsilon x^3 \left(k^2-4 x^2\right) \left[\left(k^2+4 x^2\right)^2+16 x^2\right]^{-1}}{ \sqrt{256 \varepsilon^2 x^6-\left(k^2+4 x^2\right)^2 \left[\left(k^2+4 x^2\right)^2+16 x^2\right]}}.
\end{equation}
The upper limit $\sqrt{\varepsilon^2-1}$ of integration in integral (\ref{EquatorMelBonnorSpinProbabilityIntegral}) is obtained by identifying the radius coordinate $r_0$ of closest approach with the impact parameter $b=\ell/\sqrt{\varepsilon^2-1}$. On the other hand, the lower limit $x_0$ of integration corresponds to the minimum value of $x$ for which the function $\mathcal{I}(x)$ remains real-valued.

The minimum and maximum values of the integration variable $x$ for which the sum inside the square root in the denominator of the function $\mathcal{I}(x)$ remains strictly positive depend on the parameters $\varepsilon$ and $k$. It is therefore very important before proceeding further to start by defining that region of validity that spans the parameters we are going to be using here. We do so in Fig.\,\ref{fig:MelBonnorMinimx_0Values} below. In that figure, we picked, for illustration purposes, three values of the parameter $k=B\ell$ and we plotted the ranges of the corresponding allowed values of $x$. The allowed values are contained inside the shaded areas between the continuous lines. For each case, the largest value allowed for $x$ ---\,defined by the upper branch of the continuous lines\,--- is exactly $\sqrt{\varepsilon^2-1}$. We notice that the allowed regions keep shrinking as the parameter $k$ is increasing. As $k$ increases, the lower values allowed for $x$ for each energy per unit mass $\varepsilon$ are becoming larger.
\begin{figure}[H]
    \centering
\includegraphics[scale=0.57]{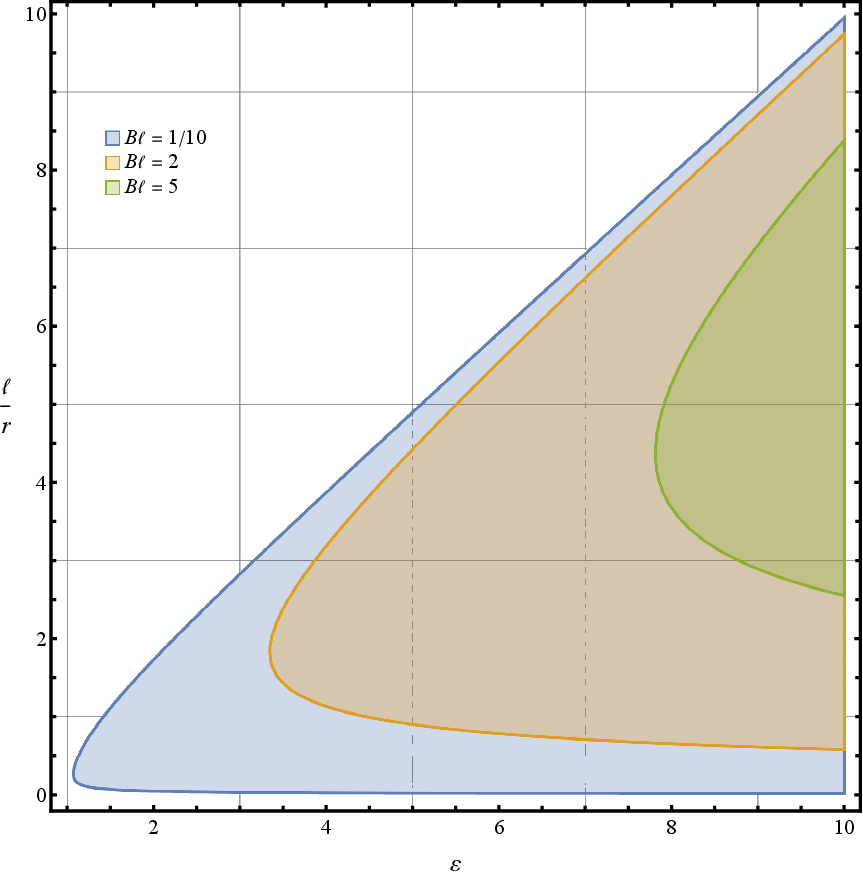}
    \caption{This plot shows the allowed values of the variable $x=\ell/r$ that make the integrand \eqref{eq:MelvinIntegrand} real-valued for each value of the energy per unit mass $\varepsilon$ of the neutrino. The allowed values for $x$ are those contained inside the shaded areas between the continuous lines. For each case, the largest value allowed for $x$ ---\,defined by the upper branch of the continuous lines\,--- is exactly $\sqrt{\varepsilon^2-1}$.}
    \label{fig:MelBonnorMinimx_0Values}
\end{figure}
Now that we have the values of $x$, we proceed to study in what follows the variations of the spin-flip probability with the variations of both $\varepsilon$ and $k$. We start by fixing the value of $\varepsilon$ in order to explore separately the two different regimes, the high-energy and the low-energy neutrinos. Once $\varepsilon$ is fixed, we vary the parameter $k$ and compute the corresponding probability $\mathcal{P}_{\ket{\nu_L}\rightarrow \ket{\nu_R}}$. After that, we let $\varepsilon$ vary from its lowest allowed possible values to very large ones by fixing the parameter $k$ to three different values chosen for illustration purposes only, without any loss of generality. 
\subsubsection{High-energy neutrinos}\label{MelBonnor High e}
Setting $\varepsilon=10^8$, we numerically evaluate the integral in Eq.\,\\(\ref{EquatorMelBonnorSpinProbabilityIntegral}) and plot the variations of $\mathcal{P}_{\ket{\nu_L}\rightarrow \ket{\nu_R}}$ against the variations of $k$. We let the latter vary from values that are arbitrary close to zero and all the way to relatively large values. The result is shown in Fig.\,\ref{fig:MelBonnorHighEnergy} below. The plot shows that the probability reaches a peak equal to one for the specific  value $0.00017$ of the product $B\ell$. This means that for a given field strength $B$, spin precession makes \textit{all} the neutrinos of specific energy per unit mass $\varepsilon=10^8$ totally flip their spin and become right-handed particles at the end of their journey provided their angular momentum per unit mass $\ell$ is about $0.00017/B$. Away from this specific value of $\ell$, the spin-flip probability of the neutrinos decreases to zero slowly for larger values of their angular momentum per unit mass $\ell$ and faster for smaller values of $\ell$.  
\begin{figure}[H]
    \centering
\includegraphics[scale=0.57]{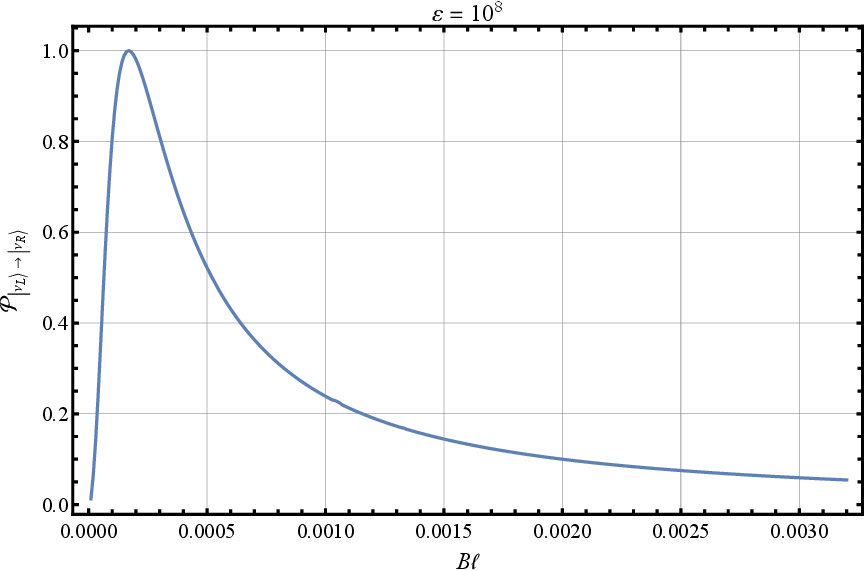}
    \caption{This plot shows the variations of spin-flip probability $\mathcal{P}_{\ket{\nu_L}\rightarrow \ket{\nu_R}}$ (given in Eq.\,(\ref{EquatorMelBonnorSpinProbabilityIntegral})) of neutrinos of energy per unit mass $\varepsilon=10^8$ as their angular momentum per unit mass $\ell$ changes relative to the electric/magnetic field strength $B$. The probability peaks at unity for the specific value $\ell= 0.00017/B$.}
\label{fig:MelBonnorHighEnergy}
\end{figure}

\subsubsection{Low-energy neutrinos}\label{MelBonnor Varying Energy}
\begin{figure}[H]
    \centering
    \includegraphics[scale=0.57]{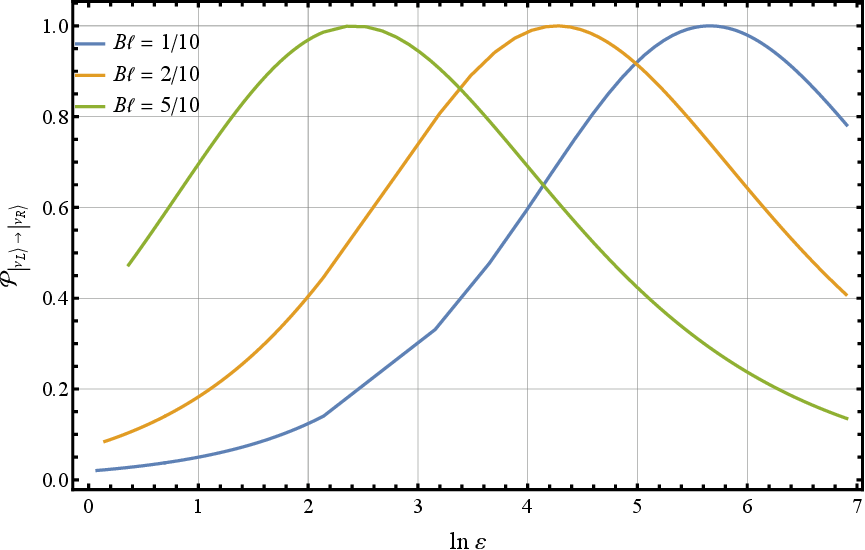}
    \caption{The variation of spin-flip probability $\mathcal{P}_{\ket{\nu_L}\rightarrow \ket{\nu_R}}$ (given in Eq.\,(\ref{EquatorMelBonnorSpinProbabilityIntegral})) with the variation of energy per unit mass $\varepsilon$ of low-energy neutrinos. The peaks in the probability corresponding to the three values chosen for the parameter $k=B\ell$ are as follows: For $k=1/10$, the maximum unit probability occurs at $\varepsilon=294$ and the minimum allowed $\varepsilon$ in this case is $\varepsilon=1.08$. For $k=2/10$, the maximum unit probability occurs at $\varepsilon=72.5$ and the minimum allowed $\varepsilon$ in this case is $\varepsilon=1.16$. For $k=5/10$, the maximum unit probability occurs at $\varepsilon=10.5$ and the minimum allowed $\varepsilon$ in this case is $\varepsilon=1.44$.}
\label{fig:MelBonnorVaryingEnergy}
\end{figure}

To investigate how the spin-flip probability changes across the vast range of possible neutrino energies, we fix now the parameter $k$ by choosing three different values for it. For each value of $k$, we let the neutrino's energy per unit mass $\varepsilon$ take on various values, ranging from its lowest possible ones to relatively large values. For convenience, in Fig.\,\ref{fig:MelBonnorVaryingEnergy} we plot those variations in terms of $\ln\varepsilon$ instead of $\varepsilon$. This has the advantage of clearly showing the different peaks the probability reaches for the three different values we chose for the parameter $k$. On the other hand, the plot also shows a neat decrease in spin-flip probability as soon as the energy per unit mass $\varepsilon$ of the neutrinos departs from the corresponding values that lead to a maximum probability in each of the three cases of $k$ chosen here.

\subsection{Spin-flip probability in the Reissner-Nordstr\"om spacetime}\label{SubSec:RN}
Before proceeding with the computation of the spin-flip probability for the Reissner-Nordstr\"om spacetime, we need to check again for which values of the angle $\theta$ does the magnitude of the angular velocity (\ref{AngularVelocityRN}) we found for spin precession in this spacetime reaches its maximum value. A straightforward calculation of the derivative of the angular velocity's magnitude $\Omega$ with respect to $\theta$ using the two nonvanishing components given in Eq.\,(\ref{AngularVelocityRN}) shows that such a derivative vanishes for $\theta=\pi/2$. As such, we conclude that, just like the case of the Melvin-Bonnor spacetime, spin precession does indeed reach its maximum angular speed when the particle propagates along the equatorial plane of the Reissner-Nordstr\"om spacetime. The effect of gravity on spin precession is, of course, the only contribution taken into account here.

Setting $\theta=\pi/2$ in the expressions (\ref{AngularVelocityRN}) yields a single nonvanishing component for the angular velocity vector that reads
\begin{equation}\label{EquatorAngularVelocityRN}
    \Omega_{\hat{2}}=\pm\frac{\varepsilon\ell}{\ell^2+r^2}.
    \end{equation}
Plugging this result into the general probability expression (\ref{GeneralProbabilityFormula}) after performing the change of variables $x=\ell/r$, we find the following probability for a left-handed neutrino to turn into a right-handed neutrino as it freely propagates and gets deflected in this spacetime:
\begin{align}\label{EquatorRNSpinProbabilityIntegral}
\mathcal{P}(\ket{\nu_L}\rightarrow \ket{\nu_R})=\sin^2\left[ \int_{0}^{\sqrt{\varepsilon^2-1}}\mathcal{I}(x)\,{\rm d}x \right],
\end{align}
where
\begin{equation}
\mathcal{I}(x)=\frac{2\varepsilon}{\left(x^2+1\right) \sqrt{\varepsilon^2-\left(x^2+1\right)\left(\mu^2 k^2x^2-\mu x+1\right)}}.
\end{equation}
We have set $\mu = r_s/\ell$ and $k=Q/r_s$. Also, for definiteness, we chose the closest approach to have the radial coordinate $r_0$ equal to twice the Schwarzschild radius $r_s$ of the black hole, a region where the large-curvature effects would be most significant. This is the origin of the upper limit of integration $\sqrt{\varepsilon^2-1}$; i.e., it corresponds to  $x=\ell/2r_s$. Concerning the lower limit of integration $x=0$, it is found by sending the radial coordinate $r$ to infinity.

In what follows, we are going to consider both high-energy and low-energy neutrinos again. First, fixing $\varepsilon$ to a large value, we will plot the variations of the probability $\mathcal{P}_{\ket{\nu_L}\rightarrow \ket{\nu_R}}$ against the variations of the parameter $k$. The latter is restricted to have the values $-1/2\leq k\leq1/2$, for it is only when the values of $k$ lie within this range that we do have a genuine black hole spacetime without a naked singularity. After that, we shall pick up two different values of $k$ and plot the variations of $\mathcal{P}_{\ket{\nu_L}\rightarrow \ket{\nu_R}}$ against the variations of $\varepsilon$. 
\subsubsection{High-energy neutrinos}\label{RN High e}
Setting $\varepsilon=10^8$, for which we have $\mu=2.5\times10^{-8}$, we numerically perform the integral in Eq.\,(\ref{EquatorRNSpinProbabilityIntegral}) and plot the variations of $\mathcal{P}_{\ket{\nu_L}\rightarrow \ket{\nu_R}}$ against the variations of $k$. The result is shown in Fig.\,\ref{fig:RNHighEnergy} below. While the probability remains extremely small and, hence, remains insignificant in the region of parameter space $-1/2\leq k\leq1/2$ and outside of it, the curve nevertheless shows a trough in the values taken by the probability that occurs for zero charge of the black hole. In other words, the spin-flip probability for high-energy neutrinos reaches its minimum only for the neutral Schwarzschild black hole, not for the charged Reissner-Nordstr\"om black hole. We have allowed the plot in that figure to extend into the region $k<-1/2$ and the region $k>1/2$ of parameter space where there is a naked singularity for the sole purpose of showing the smoothness of the curve across the two regions of the parameter space.
\begin{figure}[H]
    \centering
\includegraphics[scale=0.57]{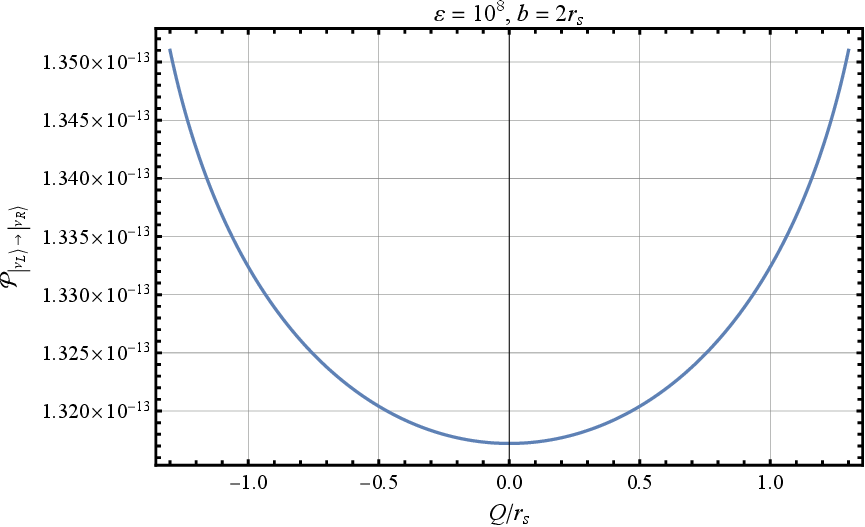}
    \caption{This graph displays the variations of the spin-flip probability (\ref{EquatorRNSpinProbabilityIntegral}) against the variations of the charge-to-mass ratio $k=Q/r_s$ of the black hole for high-energy ($\varepsilon=10^8$) neutrinos in the Reissner-Nordstr\"om black hole. The black hole region of the parameter space lies within the range $-1/2\leq k \leq 1/2$ of parameter space, outside of which the black hole exhibits a naked-singularity.}
    \label{fig:RNHighEnergy}
\end{figure}
\subsubsection{Low-energy neutrinos}\label{RN Zoomed Varying Energy}
For the case of low-energy neutrinos, we shall consider here a range of small values of $\varepsilon$ and plot the variations of $\mathcal{P}_{\ket{\nu_L}\rightarrow \ket{\nu_R}}$ against the variations of $\varepsilon$ for the two different cases of neutral ($k=0$) and charged black holes ($k\neq0$). For the latter case, we chose $k=2/20$ for the convenience of numerical calculations only, without any loss of generality. The result is shown in Fig.\,\ref{fig:RNVaryingEnergy} below. The two curves show that the probability becomes significant at low energies of the neutrinos in both cases of neutral and charged black holes. Away from the peaks, the probability slowly decreases for an increasing $\varepsilon$, but it rapidly decreases for a decreasing $\varepsilon$. In the same figure, we show a zoom-in on the region around the probability peaks for each of the two cases, neutral and charged black holes. The zoom into that region shows that the probability peaks at very slightly different values of $\varepsilon$ for neutral and charged black holes.

The remarkable fact, though, is that, as opposed to high-energy neutrinos, a value as high as unity is reached by the spin-flip probability for low-energy neutrinos, both in the charged  and in the neutral black hole cases. 
\begin{figure}[H]
    \centering
\includegraphics[scale=0.57]{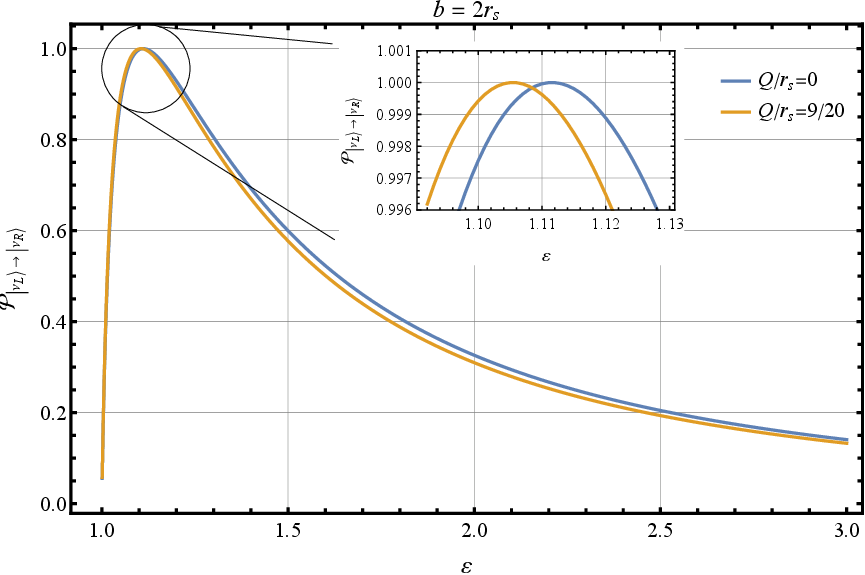}
    \caption{This graph displays the variations of the spin-flip probability (\ref{EquatorRNSpinProbabilityIntegral}) against the variations of the energy-per unit mass $\varepsilon$ of the neutrinos for both the charged Reissner-Nordstr\"om black hole (with $Q/r_s=9/20$) and the neutral Schwarzschild black hole ($Q/r_s=0$). The zoomed-in window of the graph shows the region around the peak probability for both cases, the charged and the neutral black hole. The unit probability is achieved as follows: For $Q/r_s=0$ (the Schwarzschild spacetime case), the unit probability is achieved for $\varepsilon=1.112$. The minimum energy probed for that case by our numerical computation is $\varepsilon=1.001$, for which the corresponding probability is found to be $0.055$. For $Q/r_s=9/20$, the probability reaches unity for $\varepsilon=1.105$. The minimum energy probed for this case by our numerical computation is again $\varepsilon=1.001$, for which the corresponding probability is found to be $0.058$.}
    \label{fig:RNVaryingEnergy}
\end{figure}
\subsection{Spin-flip probability in the interior Schwarzschild spacetime}\label{SubSec:InSchw}
As we did for the previous two spacetimes, before proceeding with the computation of the spin-flip probability in the interior Schwarzschild spacetime, we need to check for which values of the angle $\theta$ does the magnitude of the angular velocity (\ref{AngularVelocityInSch}) we found for spin precession reach its maximum value. Computing the derivative of the angular velocity's magnitude $\Omega$ with respect to $\theta$ using the two nonvanishing components given in Eq.\,(\ref{AngularVelocityInSch}) yields an expression that vanishes for $\theta=\pi/2$. As such, we conclude here too that spin precession does indeed reach its maximum angular speed when the particle propagates along the equatorial plane of the interior of the gravitational source.

Setting $\theta=\pi/2$ in the expressions (\ref{AngularVelocityInSch}) yields a single nonvanishing component that reads
\begin{equation}\label{EquatorAngularVelocityInSchwar}
    \Omega_{\hat{2}}=\mp \frac{2 \varepsilon\ell}{\ell ^2+r^2}\left(1-3R \sqrt{\frac{R-r_s}{R^3-r_s r^2}}\right)^{-1}.
    \end{equation}
Plugging this result into the general probability expression (\ref{GeneralProbabilityFormula}), we find the following probability for a left-handed neutrino to turn into a right-handed neutrino as it propagates and gets deflected in this spacetime:
\begin{equation}\label{InSchwarSpinProbabilityIntegral}
\mathcal{P}(\ket{\nu_L}\rightarrow \ket{\nu_R})=\sin^2\left[ \int_{1/k\mu}^{\sqrt{\varepsilon^2-1}} \mathcal{I}(x)\,{\rm d}x \right].
\end{equation}
The function $\mathcal{I}(x)$ is obtained by performing the change of variables $x=\ell/r$ and then setting $k=R/r_s$ and $\mu=r_s/\ell$. The explicit expression of the function $\mathcal{I}(x)$ is
\begin{align}
    \mathcal{I}(x)&=\frac{4 \varepsilon \left(x^2+1\right)^{-1}}{\sqrt{4 \varepsilon ^2-\left(x^2+1\right) \left(\sqrt{1-\frac{1}{k^3 \mu ^2 x^2}}-3 \sqrt{\frac{k-1}{k}}\right)^2}}
\end{align}
The upper limit $\sqrt{\varepsilon^2-1}$ of integration in integral (\ref{InSchwarSpinProbabilityIntegral}) is obtained by identifying the closest approach of the particle to the gravitational source with the impact parameter $b$ so that the largest value of $x=\ell/r$ corresponds to the smallest value of $r$, which is $b$. The lower limit of integration $1/k\mu$ is obtained by taking the largest distance of the particle from the center of the massive object to be the radius of the spherical body; i.e., we only consider the effect of gravity on spin precession of neutrinos as the latter delve inside the spherical body starting from the surface of the latter. In fact, as the interior Schwarzschild metric is only valid inside the gravitational source, the effect of the latter on spin precession can only be summed up by starting from the edge of the spherical body, outside of which is taken care of by the probability formula corresponding to the exterior Schwarzschild metric \cite{NSO}.

In what follows, we are going again to consider both high-energy and low-energy neutrinos. We start with the high-energy neutrinos by fixing the energy per unit mass $\varepsilon$ and plot the variations of the probability $\mathcal{P}_{\ket{\nu_L}\rightarrow \ket{\nu_R}}$ against the variations of the parameter $k=R/r_s$ that we allow to be as high as $10^9$. This specific value, chosen only for illustrative purposes, corresponds to the order of magnitude of Earth's size and density.

\subsubsection{High-energy neutrinos}\label{InSchwar High e}
Setting $R=10^9r_s$ and $\varepsilon=10^8$, for which we have $\mu=10^{-8}r_s/b$, we numerically evaluate the integral in Eq.\,(\ref{InSchwarSpinProbabilityIntegral}) and plot the variations of $\mathcal{P}_{\ket{\nu_L}\rightarrow \ket{\nu_R}}$ against the variations of the ratio $b/r_s$. The result is shown in Fig.\,\ref{fig:interiorSchwarzschildHighEnergy} below. We immediately notice that the probability reaches unity for a very specific value of the ratio $b/r_s$. For the particular values chosen here for $\varepsilon$ and $k$, the peak of the probability occurs at $b/r_s=10$. Away from this value of the ratio, the probability decreases to zero faster for lower values of $b/r_s$ and decreases slower for larger values of this ratio. This peak in the probability might be interpreted as a resonance that occurs in the spin-flip probability for a value of the impact parameter $b$ that is around $R/\varepsilon$ of the high-energy neutrinos. In other words, every neutrino has a different impact parameter for which it undergoes a flip in its spin with a high probability. This depends on both the energy and the size of the gravitational source.  
\begin{figure}[H]
    \centering
\includegraphics[scale=0.57]{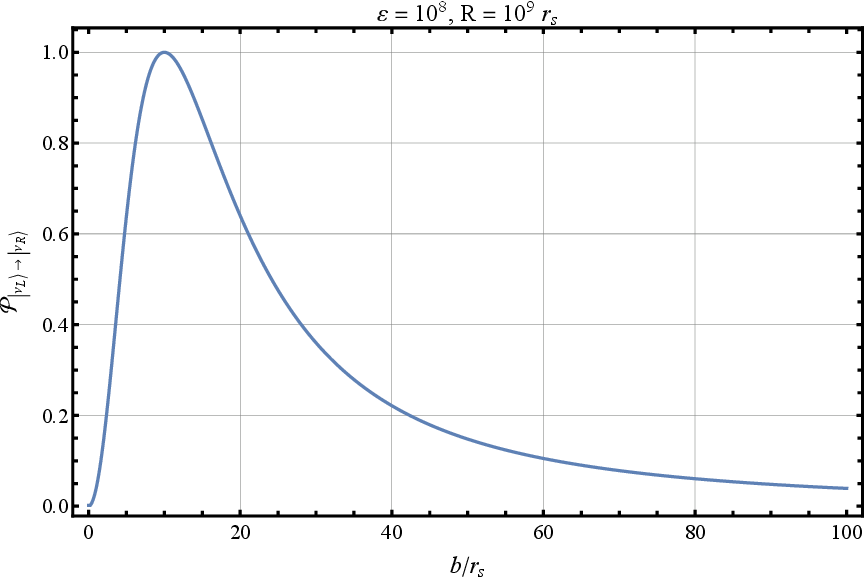}
    \caption{The curve in this graph is obtained by fixing the energy per unit mass $\varepsilon$ to be $\varepsilon=10^8$ and evaluating integral (\ref{InSchwarSpinProbabilityIntegral}) for various values of the ratio $b/r_s$. The latter is taken to span the values range going from $0$ to $100$. The parameter $k=R/r_s$ is set to $k=10^9$ which is roughly the case for planet Earth. The peak in the probability has the value one, and it occurs for $b/r_s=10$.}
\label{fig:interiorSchwarzschildHighEnergy}
\end{figure}
\subsubsection{Low-energy neutrinos}\label{InSchwar Low e}
For the case of low-energy neutrinos, we set $\varepsilon=5$ and we vary again the parameter $k=b/r_s$. For such a case, we have $\mu=0.05r_s/b$. The parameter $k=R/r_s$ is chosen again to be $10^9$. Numerically evaluating integral (\ref{InSchwarSpinProbabilityIntegral}) for the various values of $k$ yields the plot displayed in Fig.\,\ref{fig:interiorSchwarzschildLowEnergyI} below. The curve shows again a peak equal to unity for the probability  $\mathcal{P}_{\ket{\nu_L}\rightarrow \ket{\nu_R}}$, but the peak occurs when the ratio $b/r_s$ takes the very large value $1.96\times10^8$.

\begin{figure}[H]
    \centering
\includegraphics[scale=0.57]{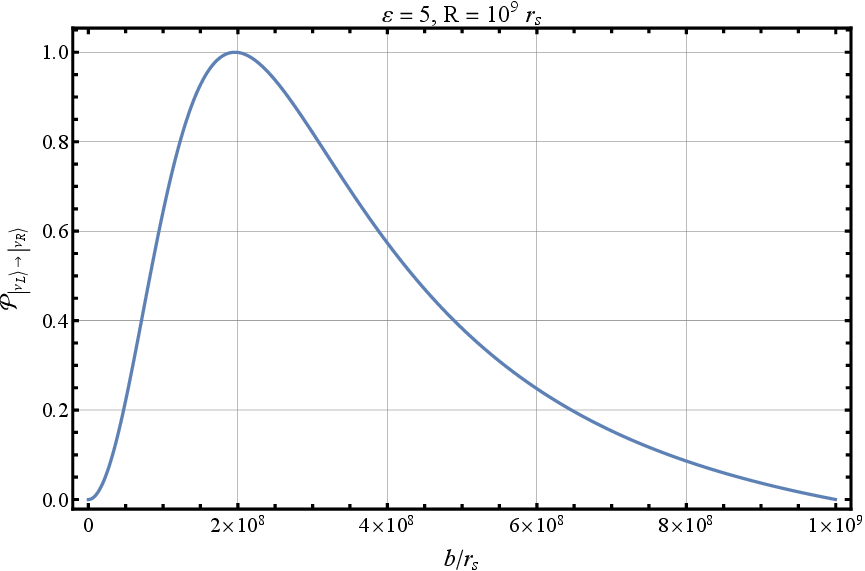}
    \caption{This graph displays the variations of the probability $\mathcal{P}_{\ket{\nu_L}\rightarrow \ket{\nu_R}}$ (given by Eq.\,(\ref{InSchwarSpinProbabilityIntegral})) against the variations of the ratio $b/r_s$ for the energy per unit mass $\varepsilon=5$ of low-energy neutrinos. The parameter $k=R/r_s$ is set to $10^9$.}
    \label{fig:interiorSchwarzschildLowEnergyI}
\end{figure}
\subsection{Spin-flip probability in the Hayward spacetime}\label{SubSec:Hayward}
The angular velocity (\ref{AngularVelocityHayward}) we found for spin precession within the Hayward spacetime is identical to the angular velocity (\ref{AngularVelocityRN}) we found for spin precession within the Reissner-Nordstr\"om spacetime. Therefore, the result that spin precession reaches its maximum along the equatorial plane, for which $\theta=\pi/2$, also applies here for the Hayward spacetime. Setting $\theta=\pi/2$ in the expressions (\ref{AngularVelocityHayward}) yields again the result (\ref{EquatorAngularVelocityRN}) we found for the single nonvanishing component of the angular velocity three-vector.
Plugging that result into the general probability expression (\ref{GeneralProbabilityFormula}) after performing the change of variables $x=\ell/r$, we find the following probability formula for a left-handed neutrino to turn into a right-handed neutrino as it propagates and gets deflected in the spacetime:
\begin{align}\label{SpinProbabilityIntegralHayward}
\mathcal{P}(\ket{\nu_L}\rightarrow \ket{\nu_R})=\sin^2\left[ \int_{0}^{\sqrt{\varepsilon^2-1}} \mathcal{I}(x)\,{\rm d}x \right],
\end{align}
where
\begin{equation}
    \mathcal{I}(x)=\frac{2\varepsilon\sqrt{k \mu ^3 x^3+3}\left(x^2+1\right)^{-1}}{ \sqrt{\varepsilon^2\left(k \mu ^3 x^3+3\right)-\left(x^2+1\right) \left(k \mu ^3 x^3-3 \mu  x+3\right)}}.
\end{equation}
We have set $k=C/r_s^2$ and $\mu=r_s/\ell$. The limits $0$ and  $\sqrt{\varepsilon^2-1}$ of integration over $x$ are obtained by sending the radial coordinate $r$ to infinity and identifying the radial coordinate with the impact parameter $b$, respectively. As we did for the case of the Reissner-Nordstr\"om spacetime, we choose here the impact parameter $b$ to be twice the Schwarzschild radius $r_s$ of the black hole, a region where the large-curvature effects would be most significant.

In what follows, we consider both high-energy and low-energy neutrinos. For high-energy neutrinos, we will plot the variations of the probability $\mathcal{P}_{\ket{\nu_L}\rightarrow \ket{\nu_R}}$ against the variations of the parameter $ k=C/r_s^2$. For low-energy neutrinos, we pick up two distinct values of the parameter $k$ and we let energy vary from the lowest possible value $\varepsilon=1$ to the value $\varepsilon=3$. 
\subsubsection{High-energy neutrinos}\label{Hayward High e}
We set here $\varepsilon=10^8$, for which we get $\mu=2.5\times10^{-8}$. We numerically evaluate integral (\ref{SpinProbabilityIntegralHayward}) for different values of the ratio $k=C/r_s^2$ and plot the variations of $\mathcal{P}_{\ket{\nu_L}\rightarrow \ket{\nu_R}}$ against the variations of $k$. The result is shown in Fig.\,\ref{fig:HaywardHighEnergy} below. The probability remains extremely small, and hence insignificant, and it increases slightly as the ratio $C/r_s^2$ increases.
\begin{figure}[H]
    \centering
\includegraphics[scale=0.57]{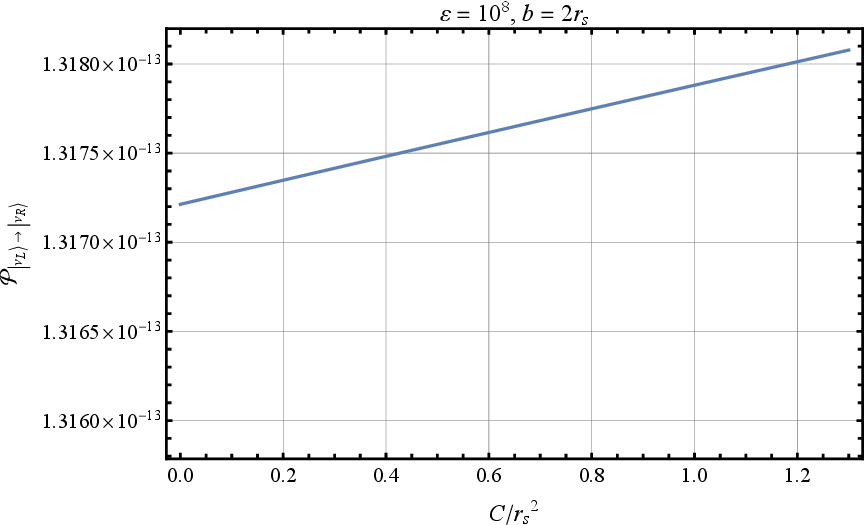}\\
    \caption{The spin-flip probability $\mathcal{P}_{\ket{\nu_L}\rightarrow \ket{\nu_R}}$ (given by Eq.\,(\ref{SpinProbabilityIntegralHayward})) as a function of the ratio $C/r_s$ in the Hayward spacetime for high-energy neutrinos, with $\varepsilon=10^8$ and an impact parameter $b=2r_s$.}
    \label{fig:HaywardHighEnergy}
\end{figure}
\subsubsection{Low-energy neutrinos}\label{Hayward Varying Energy}
\begin{figure}[H]
    \centering
\includegraphics[scale=0.57]{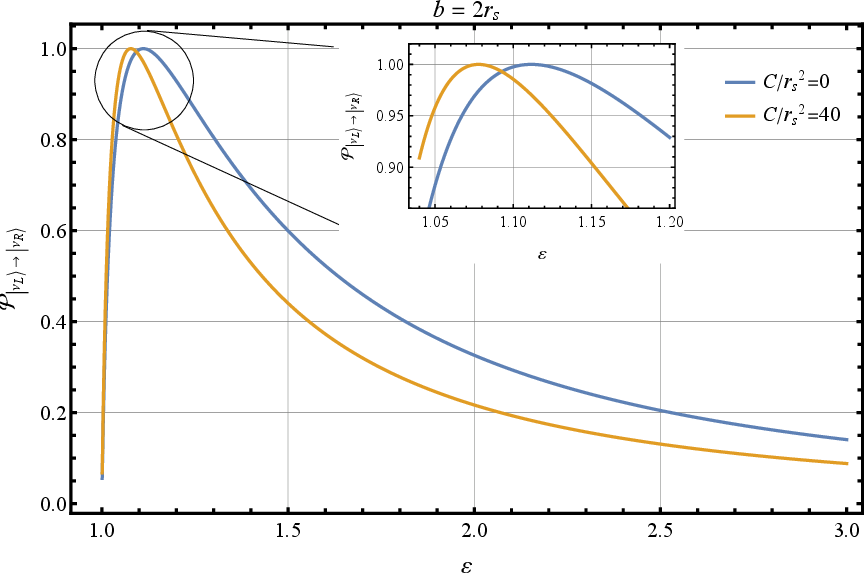}
    \caption{The spin-flip probability $\mathcal{P}_{\ket{\nu_L}\rightarrow \ket{\nu_R}}$ (given by Eq.\,(\ref{SpinProbabilityIntegralHayward})) as a function of $\varepsilon$ of low-energy neutrinos freely propagating in the Hayward spacetime. The two curves correspond to the different ratios $C/r_s^2=0$ and $C/r_s^2=40$ for the same impact parameter $b=2r_s$. The zoomed-in window of the graph shows the region around the peak probability for both cases. The unit probability is reached as follows: For $C/r_s^2=0$ (the Schwarzschild spacetime case), the unit probability is reached for $\varepsilon=1.112$. For $C/r_s^2=40$, the probability reaches unity for $\varepsilon=1.078$.}
    \label{fig:HaywardLowEnergy}
\end{figure}

In contrast to the case of high-energy neutrinos, the numerical evaluation of integral (\ref{SpinProbabilityIntegralHayward}) yields significant values for the spin-flip probability of low-energy neutrinos. Therefore, instead of considering a single value for the energy per unit mass $\varepsilon$ of such neutrinos, we investigate here the variations of the probability with the variations of $\varepsilon$ for small values of the latter. We plot the variations of $\mathcal{P}_{\ket{\nu_L}\rightarrow \ket{\nu_R}}$ against the variations of $\varepsilon$ in Fig.\,\ref{fig:HaywardLowEnergy} below for two different values of the ratio $C/r_s^2$. The case $C/r_s^2=0$ corresponds to a  neutral and singularity-free black hole that can be compared to the case of the neutral, but singular Schwarzschild black hole studied above as a special case of the more general Reissner-Nordstr\"om black hole. Comparing the blue curve of Fig.\,\ref{fig:HaywardLowEnergy} to the blue curve of Fig.\,\ref{fig:RNVaryingEnergy}, the behavior of both probabilities are the same away from their respective peaks located at $\varepsilon=1.112$. 

The case $C/r_s^2=40$ corresponds to a neutral and singula-\\
rity-free black hole embedded in an expanding universe of positive cosmological constant $C$. The corresponding curve is the yellow one. The plot clearly shows the effect of the expansion of the universe on the spin-flip probability. The latter merely shifts its peak towards lower energies of the neutrinos. The global shape of the curve remains similar to the one in the absence of cosmic expansion, only its decreasing is faster for larger values of $\varepsilon$. 

In the same figure, we show a zoom-in on the region around the probability peaks for each of the two cases, in the presence and in the absence of cosmic expansion. The zoom into that region neatly displays the separate peaks in the probability. The two peaks lie at very slightly different values of $\varepsilon$, the latter being at a smaller value in the presence of cosmic expansion.
\subsection{Spin-flip probability in the Bardeen spacetime}\label{SubSec:Bardeen}
The angular velocity (\ref{AngularVelocityBardeen}) we found for spin precession within the Bardeen spacetime is identical to the angular velocity we found for spin precession within the Reissner-Nordstr\"om and the Hayward spacetimes. As argued above, for such an angular velocity, spin precession reaches its maximum along the equatorial plane, for which $\theta=\pi/2$. The effect of gravity on spin precession is, of course, the only contribution taken into account here too.

Setting $\theta=\pi/2$ in the expressions (\ref{AngularVelocityBardeen}) yields again the result (\ref{EquatorAngularVelocityRN}) we found for the single nonvanishing component of the angular velocity vector. Plugging that result into the general probability expression (\ref{GeneralProbabilityFormula}) after performing the change of variables $x=\ell/r$ and setting $k=q/r_s$, we find the following probability for a left-handed neutrino to turn into a right-handed neutrino as it propagates and gets deflected in the spacetime:
\begin{equation}\label{SpinProbabilityIntegralBardeen}
\mathcal{P}(\ket{\nu_L}\rightarrow \ket{\nu_R})=\sin^2\left[ \int_{0}^{\sqrt{\varepsilon^2-1}} \mathcal{I}(x)\,{\rm d}x \right],
\end{equation}
where
\begin{equation}
    \mathcal{I}(x)=\frac{2\varepsilon\left(k^2 \mu^2 x^2+1\right)^{3/2}\left(x^2+1\right)^{-1}}{\sqrt{\mu x^3+\mu x+\left(\varepsilon^2-x^2-1\right)\left(k^2 \mu^2 x^2+1\right)^{3/2}}}.
\end{equation}
We have set $\mu=r_s/\ell$. As we did for the Hayward spacetime case above, the limits $0$ and  $\sqrt{\varepsilon^2-1}$ of integration over $x$ are obtained by sending the radial coordinate $r$ to infinity and identifying the radial coordinate with the impact parameter $b$, respectively. Also, we choose here the closest approach coordinate radius $r_0=b$ to be again twice the Schwarzschild radius $r_s$ of the black hole, a region where the large-curvature effects would be most significant.

In what follows, we are going to consider both high-energy and low-energy neutrinos again, and start by fixing $\varepsilon$ for which we will plot the variations of the probability $\mathcal{P}_{\ket{\nu_L}\rightarrow \ket{\nu_R}}$ against the variations of the parameter $k=q/r_s$. We let the latter take values within the range $-10\leq k\leq10$ for illustration purposes. We then vary the energy per unit mass $\varepsilon$ and choose three different values of the ratio $q/r_s$.  
\subsubsection{High-energy neutrinos}\label{Bardeen High e}
We set $\varepsilon=10^8$, which gives $\mu=2.5\times10^{-8}$. We numerically evaluate integral (\ref{SpinProbabilityIntegralBardeen}) for different values of the ratio $k=q/r_s^2$ and plot the variations of $\mathcal{P}_{\ket{\nu_L}\rightarrow \ket{\nu_R}}$ against the variations of $k$. The result is shown in Fig.\,\ref{fig:BardeenHighEnergy} below. The probability remains extremely small, and hence insignificant again, and it increases slightly as the ratio $q/r_s^2$ departs from zero.
\begin{figure}[H]
    \centering
    \includegraphics[scale=0.57]{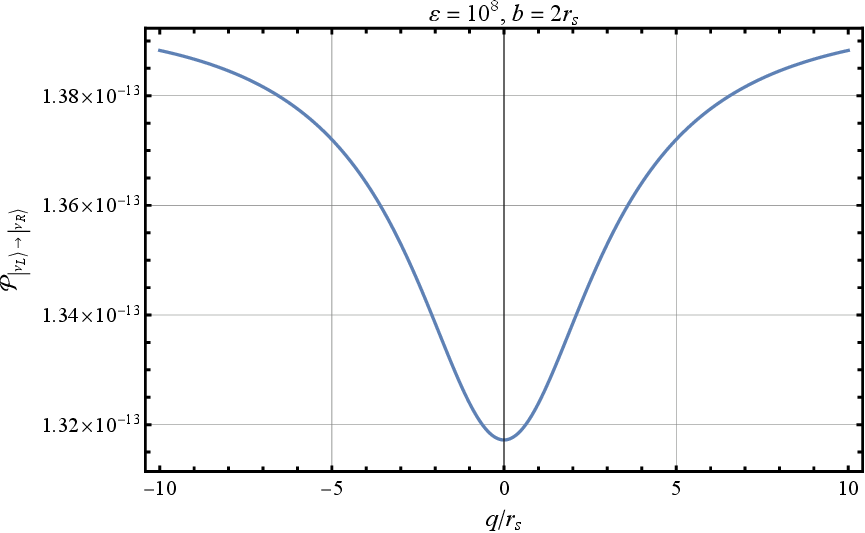}
    \caption{The variations of the spin-flip probability  $\mathcal{P}_{\ket{\nu_L}\rightarrow \ket{\nu_R}}$ (given by Eq.\,(\ref{SpinProbabilityIntegralBardeen})) for high-energy neutrinos ($\varepsilon=10^8$) as the ratio $q/r_s$ departs from zero within the Bardeen spacetime. The impact parameter $b$ has been chosen to be $2r_s$.}
    \label{fig:BardeenHighEnergy}
\end{figure}
\subsubsection{Low-energy neutrinos}\label{Bardeen Varying Energy}
Unlike the case of high-energy neutrinos, the spin-flip probability becomes considerably more significant for low-energy neutrinos. Indeed, when the energy per unit mass $\varepsilon$ of the neutrinos is much smaller, the numerical evaluation of integral (\ref{SpinProbabilityIntegralBardeen}) gives values for the spin-flip probability as high as unity. We are therefore going to consider here a continuous range of possible small values of $\varepsilon$ and plot the corresponding variations of the spin-flip probability. We limit the range of $\varepsilon$ to be below $\varepsilon=3$ as the peak in probability is reached closer to $\varepsilon=1$ as shown in Fig.\,\ref{fig:BardeenLowEnergy} below. We consider three different values of the ratio $q/r_s^2$. The case $q/r_s^2=0$, plotted with a blue curve, corresponds to the Schwarzschild black hole. The two other cases, $q/r_s=2$ and $q/r_s=3$, plotted in orange and green, respectively, correspond to black holes carrying a magnetic charge. Comparing the blue curve to the other two, we see that the behavior of the probability consists in shifting its peak towards even lower values of $\varepsilon$. For all three curves, the peak is reached around the value $\varepsilon=1.1$.
\begin{figure}[H]
    \centering
    \includegraphics[scale=0.57]{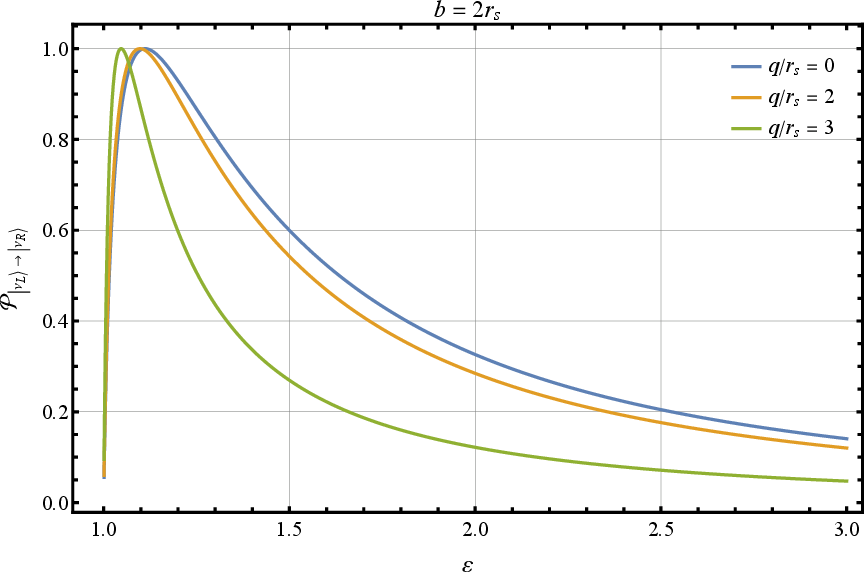}
    \caption{This graph displays the variations of the spin-flip probability (\ref{SpinProbabilityIntegralBardeen}) with the variations of the energy per unit mass $\varepsilon$ for three different values of the ratio $k=q/r_s$. The maximum probability of value one is reached for each of these three different values as follows: For $q/r_s=0$ (the Schwarzschild spacetime case), the maximum probability is reached for $\varepsilon=1.112$. For $q/r_s=2$,  the maximum probability is reached for $\varepsilon=1.098$. For $q/r_s=3$,  the maximum probability is reached for $\varepsilon=1.047$.}
\label{fig:BardeenLowEnergy}
\end{figure}
\subsection{Spin-flip probability in the Kiselev spacetime}\label{SubSec:Kiselev}
The Kiselev spacetime induces the same angular velocity for spin precession as the one induced by the Bardeen spacetime we examined above. Therefore, we shall also study here the spin-flip probability of particles moving along the equatorial plane at $\theta=\pi/2$ for which the angular velocity of spin precession is maximal. Setting $\theta=\pi/2$ in the expressions (\ref{AngularVelocityKiselev}) yields again the result (\ref{EquatorAngularVelocityRN}) we found for the single nonvanishing component of the angular velocity vector in the Reissner-Nordstr\"om spacetime. Plugging that result into the general probability expression (\ref{GeneralProbabilityFormula}) after performing the change of variables $x=\ell/r$ and setting $k=r_q/r_s$, we find the following probability for a left-handed neutrino to turn into a right-handed neutrino as it propagates and gets deflected in the spacetime:
\begin{equation}\label{SpinProbabilityIntegralKiselev}
        \mathcal{P}(\ket{\nu_L}\rightarrow \ket{\nu_R})=\sin^2\left[ \int_{x_{\rm CH}}^{\sqrt{\varepsilon^2-1}} \,{\rm d}x \right]
        \end{equation}
where
\begin{equation}
    \mathcal{I}(x)=\frac{2\varepsilon\left(x^2+1\right)^{-1}}{ \sqrt{\varepsilon^2+\left(x^2+1\right) \left[(k \mu x)^{3 \omega_q+1}+\mu x-1\right]}}.
\end{equation}
We have set $\mu=r_s/\ell$. The limits $x_{\rm CH}$ and  $\sqrt{\varepsilon^2-1}$ of integration over $x$ are obtained as follows. The upper limit comes from the closest approach coordinate radius $r_0=b$, the impact parameter $b$ being identified with twice the Schw-\\
arzschild radius $r_s$ of the black hole, a region where the large-curvature effects would be most significant. The lower limit $x_{\rm CH}$, where ``CH'' stands for cosmological horizon, is obtained by identifying the largest coordinate radius $r$ with the radial location of the cosmological horizon of the spacetime. Such a location is the largest root one obtains for the equation $\Lambda=0$, where  $\Lambda$ is the metric component in Eq.\,(\ref{KiselevMetric}). As our goal is, among other things, to study the variations of the spin-flip probability with the variations of the parameter $\omega_q$, and as $\Lambda$ depends on $r_s$, $r_q$ and $\omega_q$, the lower integration limit $x_{\rm CH}$ needs to be recalculated for each new value assigned to the parameter $\omega_q$. The Mathematica code we built to perform such a calculation for various values of $\omega_q$ thus computes the new location of the cosmological horizon for every new value it picks up for $\omega_q$ inside the selected range. 

In what follows, we separately consider high-energy and low-energy neutrinos. We start with the case of high-energy neutrinos by fixing $\varepsilon$ and plot the variations of the probability $\mathcal{P}_{\ket{\nu_L}\rightarrow \ket{\nu_R}}$ against the variations of the quintessence parameter $\omega_q$. For low-energy neutrinos, we let $\varepsilon$ vary from the lowest possible value $\varepsilon=1$ to $\varepsilon=2.5$ by fixing the parameter $\omega_q$ to two different values.  
\subsubsection{High-energy neutrinos}\label{Kiselev High e}
We set $\varepsilon=10^8$, for which the parameter $\mu$ takes the value $\mu=2.5\times10^{-8}$. We fix the ratio $r_q/r_s$ to be as high as $10^5$. We let the quintessence parameter vary from $\omega_q=-1$ to $\omega_q>-0.67$. Evaluating numerically integral (\ref{SpinProbabilityIntegralKiselev}) for the different values of the parameter $\omega_q$ and plotting the corresponding variations of $\mathcal{P}_{\ket{\nu_L}\rightarrow \ket{\nu_R}}$, we obtain the result shown in Fig.\,\ref{fig:KiselevHighEnergy} below. The probability remains very small and it decreases as the parameter $\omega_q$ increases and departs from the value $-1$ which corresponds to the de Sitter spacetime. 
\begin{figure}[H]
    \centering
\includegraphics[width=1\linewidth]{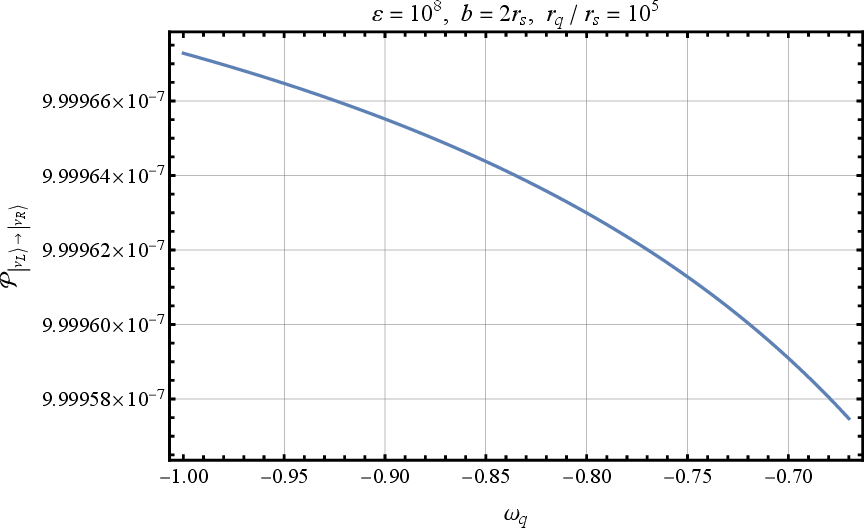}
    \caption{Variations of spin-flip probability (\ref{SpinProbabilityIntegralKiselev}) for high-energy neutrinos, with $\varepsilon=10^8$. We chose $b=2r_s$ and $r_q/r_s=10^5$. The quintessence parameter $\omega_q$ is varied from $-1$ to $-0.67$.}
    \label{fig:KiselevHighEnergy}
\end{figure}

\begin{figure}[H]
    \centering
    \includegraphics[width=1\linewidth]{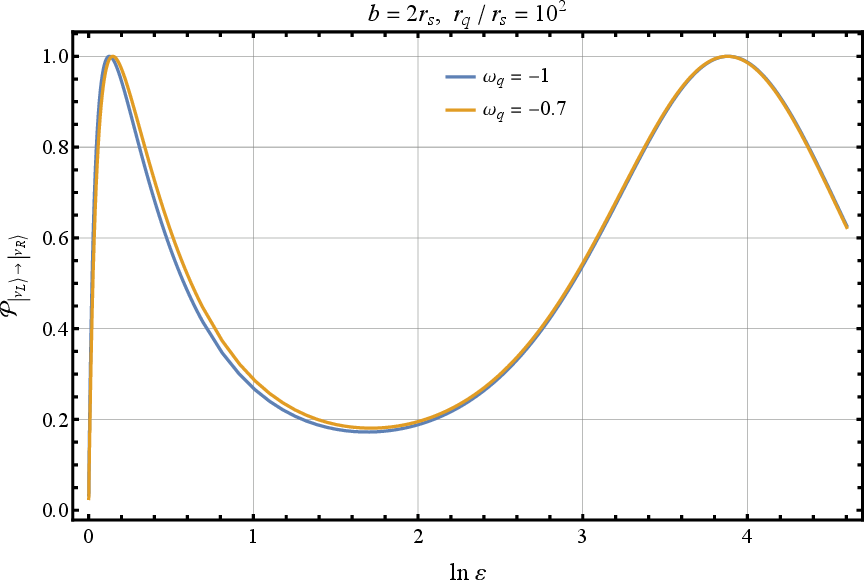}
    \caption{Variations of the spin-flip probability (given by Eq.\,(\ref{SpinProbabilityIntegralKiselev})) for different values of the energy per unit mass $\varepsilon$ of low-energy neutrinos. The two curves are obtained for $b=2r_s$, $r_q/r_s=10^2$ and for two different values of $\omega_q$. The probability peaks in each case as follows: For $\omega_q=-1$, which corresponds to the Schwarzschild-de Sitter spacetime, the peak values equal unity and they are reached for $\varepsilon=1.14$ and $\varepsilon=48.5$ while the value of the relative minimum in between is $0.172$ and is reached for $\varepsilon=5.50$. For $\omega_q=-0.7$, the peak values equal unity again and they are reached for $\varepsilon=1.16$ and $\varepsilon=48.3$ while the value of the relative minimum is $0.180$ and is reached for $\varepsilon=5.50$.}
    \label{fig:KiselevLowEnergy}
\end{figure}

\subsubsection{Low-energy neutrinos}\label{Kiselev Varying Energy}
As with the previous spacetimes, low-energy neutrinos experience spin precession more than the high-energy neutrinos do, for the spin-flip probability $\mathcal{P}_{\ket{\nu_L}\rightarrow \ket{\nu_R}}$ one extracts from formula (\ref{SpinProbabilityIntegralKiselev}) reaches unity. For this reason, instead of fixing the energy per unit mass $\varepsilon$, we are going to vary the latter from its lowest possible value to $\varepsilon=100$. By doing so, we fix the quintessence parameter to two different values $\omega_q=-1$ and $\omega_q=-0.7$. The numerical evaluation of integral (\ref{SpinProbabilityIntegralKiselev}) then yields the two curves displayed in Fig.\,\ref{fig:KiselevLowEnergy}. In each case, the spin-flip probability exhibits two peaks and one trough. After the second peak, the probability decreases for larger values of $\varepsilon$. The values of $\varepsilon$ corresponding to the two probability peaks and the relative minimum are given in the caption of Fig.\ref{fig:KiselevLowEnergy}.

\subsection{Spin-flip probability in the Schwarzschild-Melvin spacetime}\label{SubSec:SchwarMel}
Before proceeding with the computation of the spin-flip probability for the Schwarzschild-de Sitter spacetime, we need to check for which values of the angle $\theta$ does the magnitude of the angular velocity (\ref{AngularVelocitySchwarMel}) we found for spin precession in this spacetime reach its maximum value. A straightforward calculation of the derivative of the angular velocity's magnitude $\Omega$ with respect to $\theta$ using the two nonvanishing components given in Eq.\,(\ref{AngularVelocitySchwarMel}) reveals that such a derivative vanishes for $\theta=\pi/2$. As such, we conclude that, just like the other spacetimes we dealt with above, spin precession does indeed reach its maximum angular speed when the particle propagates along the equatorial plane of the Schwarzschild-Melvin spacetime when only the contribution of gravity is taken into account. 

Setting $\theta=\pi/2$ in the expressions (\ref{AngularVelocitySchwarMel}) yields a single nonvanishing component for the angular velocity vector that reads
\begin{equation}\label{EquatorAngularVelocitySchwarMel}
    \Omega_{\hat{2}}=\mp\frac{64 \varepsilon  \left(B^2 r^2-4\right) \ell }{\left(B^2 r^2+4\right)^2 \left[16 r^2+\left(B^2 r^2+4\right)^2 \ell ^2\right]}.
    \end{equation}
Plugging this result into the general probability formula (\ref{GeneralProbabilityFormula}), we find the following probability for a left-handed neutrino to turn into a right-handed neutrino as it propagates and gets deflected in the spacetime:
\begin{equation}\label{SpinProbabilityIntegralSchwarMel}
    \mathcal{P}(\ket{\nu_L}\rightarrow \ket{\nu_R})=\sin^2\left[ \int_{x_0}^{\sqrt{\varepsilon^2-1}} \mathcal{I}(x)\,{\rm d}x \right],
\end{equation}
where the function $\mathcal{I}(x)$ is obtained after performing the change of variables $x=\ell/r$ and setting $k=B\ell$ and $\mu=r_s/\ell$, and it reads
\begin{align}
    &\mathcal{I}(x)=\nonumber\\
    &\frac{128 \varepsilon x^3 \left(k^2-4 x^2\right) \left[\left(k^2+4 x^2\right)^2+16 x^2\right]^{-1}}{ \sqrt{256 \varepsilon^2 x^6+(\mu x-1)\left(k^2+4 x^2\right)^2 \left[\left(k^2+4 x^2\right)^2+16 x^2\right] }}.
\end{align}

The limits $x_0$ and  $\sqrt{\varepsilon^2-1}$ of integration over $x$ are obtained as follows. The upper limit comes from the closest approach coordinate radius $r_0=b$; the impact parameter $b$ being identified with twice the Schwarzschild radius $r_s$ of the black hole, a region where the large-curvature effects would be most significant. The lower limit $x_0$ is obtained by identifying the largest coordinate radius $r$ with the smallest radial coordinate for which the sum inside the square root in the denominator of the function $\mathcal{I}(x)$ is positive. Such a value needs to be computed for each value of the parameters $\varepsilon$, $k=B\ell$ and $\mu=r_s\ell$. As we evaluate the integral using Mathematica, our code has been developed to evaluate the integral after it performs such a calculation of the lower limit for every value considered for those three parameters.

We consider separately in what follows high-energy and low-energy neutrinos. We start with the case of high-energy neutrinos by fixing $\varepsilon$ and plot the variations of the probability $\mathcal{P}_{\ket{\nu_L}\rightarrow \ket{\nu_R}}$ against the variations of the parameter $k$. We then proceed to study the variations of the probability for low-energy neutrinos by letting their energy per unit mass $\varepsilon$ vary from the lowest possible value $\varepsilon=1$ to $\varepsilon=e^7$ for three different values of the parameter $k$.
\subsubsection{High-energy neutrinos}\label{SchwarMel High e}
We set $\varepsilon=10^8$, for which the parameter $\mu$ takes the value $\mu=2.5\times10^{-8}$. We fix the impact parameter to $b=2r_s$. On the other hand, we let the parameter $k=B\ell$ vary from $0$ to $0.003$. The numerical evaluation of integral (\ref{SpinProbabilityIntegralSchwarMel}) for the different values of the parameter $B\ell$ gives the result shown in Fig.\,\ref{fig:SchwarzschildMelvinHighEnergy} below. The probability peaks at unity for $k=0.00017$ and decreases away from that value of $k$.

Among all of the spacetimes we considered up to now, the Schwarzschild-Melvin spacetime is so far the only one that allows large spin-flip probabilities for high-energy neutrinos. 

\begin{figure}[H]
    \centering
    \includegraphics[width=1\linewidth]{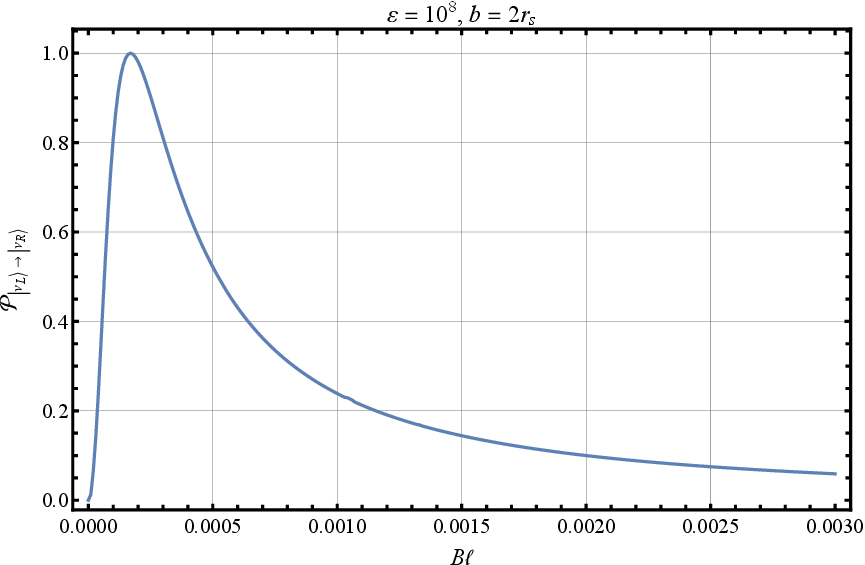}
    \caption{Spin-flip probability (as given by Eq.\,(\ref{SpinProbabilityIntegralSchwarMel})) for high-energy neutrinos ($\varepsilon=10^8$) for various values of the product $B\ell$. The peak in the probability is one, and it is reached for $B\ell=0.00017$.}
    \label{fig:SchwarzschildMelvinHighEnergy}
\end{figure}
\subsubsection{Low-energy neutrinos}\label{SchwarMel Varying Energy}
We consider now low-energy neutrinos and study their spin-flip probability. We are going to vary the energy per unit mass of the neutrinos from the lowest possible value $\varepsilon=1$ to $\varepsilon=e^7$. We keep the impact parameter $b=2r_s$, but we consider now three different values of the parameter $k=B\ell$. The numerical evaluation of integral (\ref{SpinProbabilityIntegralSchwarMel}) then yields the two curves displayed in Fig.\,\ref{fig:SchwarzschildMelvinLowEnergy} below. The remarkable thing is that spin-flip probability neatly displays again two different peaks rather than just one. In between each of those peaks, the probability drops to a relative minimum. Away from those peaks and relative minima, the probability in all three cases decreases to zero, faster for smaller values of $\varepsilon$ and slower for larger values of $\varepsilon$. The values of $\varepsilon$ corresponding to those peaks and minima are given in the caption of Fig.\,\ref{fig:SchwarzschildMelvinLowEnergy} below.
\begin{figure}[H]
    \centering
    \includegraphics[width=1\linewidth]{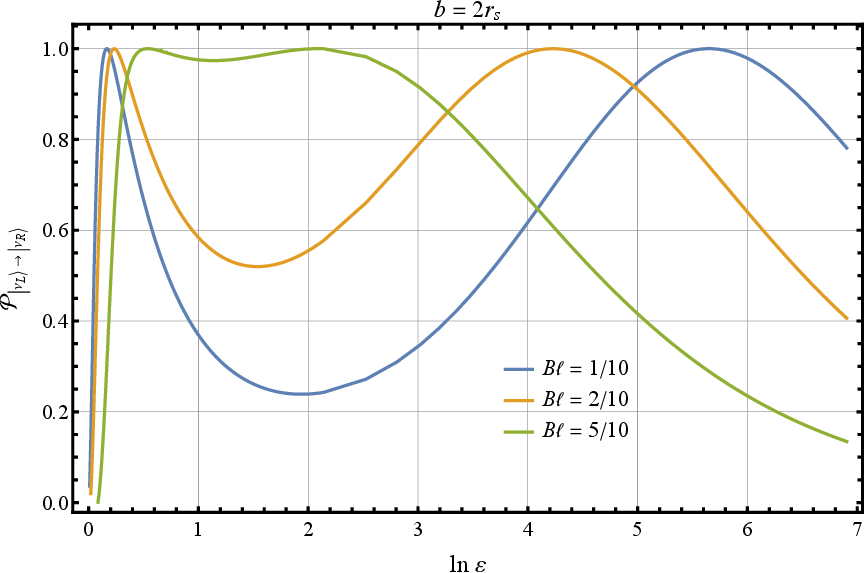}
    \caption{The three curves in this plot represent the spin-flip probability (\ref{SpinProbabilityIntegralSchwarMel}) for low-energy neutrinos with various energies, which, for convenience, are taken with their logarithmic values. The impact parameter is fixed to $b=2r_s$, whereas the parameter $B\ell$ is chosen to take on three different fixed values. The peaks and the relative minima for each case are as follows: For $B\ell=1/10$, the two probability peaks have the value one and are reached for $\varepsilon=1.18$ and $\varepsilon=285$, respectively. The relative minimum of probability for this case has the value $0.24$ and it is reached for $\varepsilon=6.95$. For $B\ell=2/10$, the two probability peaks have the value one and are reached for $\varepsilon=1.26$ and $\varepsilon=68.5$, respectively. The relative minimum of probability for this case has the value $0.52$ and it is reached for $\varepsilon=4.66$. For $B\ell=5/10$, the two probability peaks have also the value one and are reached for $\varepsilon=1.71$ and $\varepsilon=8.09$, respectively. The relative minimum of probability for this case has the value $0.97$ and it is reached for $\varepsilon=3.12$.}
\label{fig:SchwarzschildMelvinLowEnergy}
\end{figure}


\subsection{Spin-flip probability in the Weyl spacetime}\label{SubSec:Weyl}
As the angular velocity of spin precession within the Weyl spacetime is different from any of the angular velocities found for the previous spacetimes, we need to verify for which values of the angle $\theta$ does the magnitude of the angular velocity (\ref{AngularVelocityWeylMetric}) we found for spin precession in this spacetime reach its maximum value. A straightforward calculation of the derivative of the angular velocity's magnitude $\Omega$ with respect to $\theta$ using the three nonvanishing components given in Eq.\,(\ref{AngularVelocityWeylMetric}) shows that such a derivative vanishes for $\theta=\pi/2$. As such, we conclude that, just like the other spacetimes we dealt with above, spin precession does indeed reach its maximum angular speed when the particle propagates along the equatorial plane of the Weyl spacetime. 

Setting $\theta=\pi/2$ in the expressions (\ref{AngularVelocitySchwarMel}) yields a single nonvanishing component for the angular velocity vector that reads
\begin{equation}\label{EquatorAngularVelocityWeyl}
    \Omega_{\hat{2}}=\mp\frac{\varepsilon  (r_s-2 r) \ell  \exp \left(\frac{4r_s}{8 r}+\frac{r_s^2}{8 r^2}\right)}{2 r \left[\exp \left(\frac{r_s}{r}\right) r^2+\ell ^2\right]}.
    \end{equation}
Plugging this result into the general probability expression (\ref{GeneralProbabilityFormula}) after performing the change of variables $x=\ell/r$, we find the following probability for a left-handed neutrino to turn into a right-handed neutrino as it propagates and gets deflected in the spacetime:
\begin{equation}\label{SpinProbabilityIntegralWeyl}
\mathcal{P}(\ket{\nu_L}\rightarrow \ket{\nu_R})=\sin^2\left[ \int_{0}^{\sqrt{\varepsilon^2-1}} \mathcal{I}(x)\,{\rm d}x \right],
\end{equation}
where
\begin{equation}
    \mathcal{I}(x)=\frac{\varepsilon (\mu  x-2) \exp \left(\frac{3}{2}\mu x\right)}{\left[\exp (x \mu )+x^2\right] \sqrt{\varepsilon^2 \exp (2 x \mu )-\exp (x \mu )-x^2}}.
\end{equation}
We have set $\mu=r_s/\ell$. The limit $\sqrt{\varepsilon^2-1}$ of integration over $x$ is obtained by setting the closest approach $r_0=b$, the impact parameter $b$ being considered very close to the Schwar-\\
zschild radius $r_s$ of the black hole, a region where the large-curvature effects would be most significant. The lower limit $0$ is obtained by sending the largest coordinate radius $r$ to infinity.

In what follows, we consider only low-energy neutrinos. The reason is that the probability formula (\ref{SpinProbabilityIntegralWeyl}) displays only one parameter besides the energy per unit mass $\varepsilon$ of the neutrinos, which is the impact parameter of the latter. On the other hand, a numerical evaluation of integral (\ref{SpinProbabilityIntegralWeyl}) for high-energy neutrinos yields extremely small, and hence insignificant values for the spin-flip probability for any value of the impact parameter $b$. For low-energy neutrinos, however, the probability is found to be considerably high and reaches even unity. For low-energy neutrinos, we are going therefore to let $\varepsilon$ vary from its lowest possible value $\varepsilon=1$ to $\varepsilon=6$ by fixing the impact parameter $b$ to three different values.
\subsubsection{Low-energy neutrinos}\label{Weyl Varying Energy}
This spacetime induces a large spin-flip probability for low-energy neutrinos unlike what high-energy neutrinos experience. By varying the energy per unit mass of the neutrinos and plotting the variations of the probability for three different values of the impact parameter $b$ ($2r_s$, $3r_s$ and $8r_s$), we obtain the results displayed in Fig.\,\ref{fig:WeylLowEnergy} below. Spin-flip probability neatly peaks around values of $\varepsilon$ not very far from $\varepsilon=1$. In both cases, the probability decreases from unity, faster for lower values of $\varepsilon$ and slower for larger values of $\varepsilon$. The decrease in the probability from its peak occurs for larger values of the impact parameter $b$.
\begin{figure}[H]
    \centering
    \includegraphics[scale=0.57]{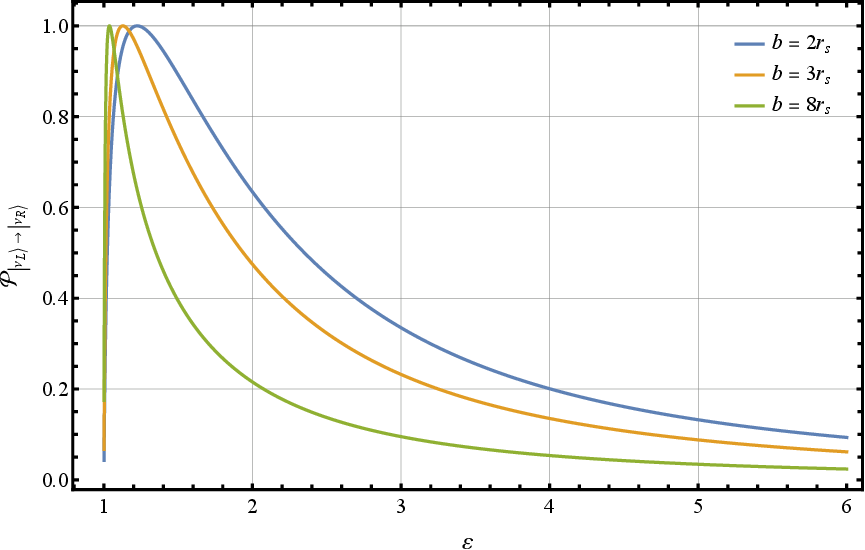}
    \caption{Spin-flip probability (given by Eq.\,(\ref{SpinProbabilityIntegralWeyl})) for various values of the energy per unit mass $\varepsilon$ of high-energy neutrinos within the Weyl spacetime. The impact parameter $b$ is fixed to three different values. For each value, we let the energy per unit mass $\varepsilon$ of the neutrino vary from $\varepsilon=1$ to $\varepsilon=6$. The peaks in the probability for each case reach unity and are obtained as follows: For $b/r_s=2$, the maximum probability is reached for $\varepsilon=1.225$. For $b/r_s=3$, the maximum probability is reached for $\varepsilon=1.128$. For $b/r_s=8$, the maximum probability is reached for $\varepsilon=1.037$.}
    \label{fig:WeylLowEnergy}
\end{figure}
\subsection{Spin-flip probability in the Kerr spacetime}\label{SubSec:NeutrinoInKerr}

As the angular velocity of spin precession within the Kerr spacetime is rather much more involved than any of the expressions we obtained so far, we have displayed the result explicitly in Appendix \ref{Sec:AppB}. However, it is not difficult to see that the values of the angle $\theta$ for which the spacetime dragging effect, and hence spin precession, is maximal along the equatorial plane of the gravitational source.

Setting $\theta=\pi/2$ in the expressions (\ref{AppBOmega1}) and (\ref{AppBOmega2}) yields the single nonvanishing component (\ref{Pi/2AngularVelocityKerr}) for the angular velocity vector. 
Plugging that result into the general probability expression (\ref{GeneralProbabilityFormula}), and making the change of variables $x=\ell/r$, we find the following probability for a left-handed neutrino to turn into a right-handed neutrino as it propagates and gets deflected in the spacetime:
\begin{equation}\label{SpinProbabilityIntegralKerr}
\mathcal{P}(\ket{\nu_L}\rightarrow \ket{\nu_R})=\sin^2\left[ \int_{0}^{\sqrt{\varepsilon^2-1}} \frac{\mathcal{I}(x)}{\mathcal{J}(x)}\,{\rm d}x \right],
\end{equation}
where the functions $\mathcal{I}(x)$ and $\mathcal{J}(x)$, obtained after setting $k=a/r_s$, read
\begin{align}
    \mathcal{I}(x)&=k \mu ^2 x \Big{[}1+x \left(k^2 \mu ^2 x-\mu  \left[x^2+1\right]+x\right)\nonumber\\
    &\quad-\varepsilon^2 \left(\mu  x \left[k^2 \mu  x (\mu  x+2)-2\right]+4\right)\Big{]}\nonumber\\
    &\quad+\varepsilon \left[-2+\mu  x\left(4+ \mu  x\left[-2+k^2 \mu  x\left(1+2 \mu  x\right)\right]\right)\right],
    \end{align}
    and
    \begin{align}
    \mathcal{J}(x)&=\Big{[}1+\mu^2 x^4\left(\varepsilon k\mu-1\right)^2+\mu x^3\left(k\mu[2\varepsilon-k\mu]-2\right)\nonumber\\
&\quad+\mu^2x^2\left(k^2+1\right)+x^2-2\mu x\Big{]}\nonumber\\
&\quad\times\Big{[}\varepsilon^2-1+\varepsilon^2k^2\mu^2x^2(\mu x+1)-2\varepsilon k\mu^2 x^3\nonumber\\
&\quad+x^2[\mu(x-k^2\mu)-1]+\mu x\Big{]}.
\end{align}
The limit $\sqrt{\varepsilon^2-1}$ of integration over $x$ is obtained by setting the closest approach $r_0=b$, the impact parameter $b$ being considered very close to the Schwarzschild radius $r_s$ of the black hole, a region where the large-curvature effects would be most significant. The lower limit $0$ is obtained by sending the largest coordinate radius $r$ to infinity.

We will now consider separately in what follows high-energy and low-energy neutrinos. We start with the case of high-energy neutrinos by fixing $\varepsilon$ and plot the variations of the probability $\mathcal{P}_{\ket{\nu_L}\rightarrow \ket{\nu_R}}$ against the variations of the parameter $k=a/r_s$. We then proceed to study the variations of the probability for low-energy neutrinos by letting their energy per unit mass $\varepsilon$ vary from the lowest possible value $\varepsilon=1$ to $\varepsilon=3$ for two different values of the parameter $k$.
\subsubsection{High-energy neutrinos}\label{RN High e}
Setting $\varepsilon=10^8$, for which we have $\mu=2.5\times10^{-8}$, we numerically evaluate the integral in Eq.\,(\ref{SpinProbabilityIntegralKerr}) and plot the variations of $\mathcal{P}_{\ket{\nu_L}\rightarrow \ket{\nu_R}}$ against the variations of the parameter $k$. The result is shown in Fig.\,\ref{fig:KerrHighEnergy} below. While the probability remains extremely small, and hence insignificant, we nevertheless interestingly see a peak and a rise in the probability for a specific value of the rotation of the spinning black hole. After that, the probability decreases as the ratio $a/r_s$ increases.
\begin{figure}[H]
    \centering
    \includegraphics[scale=0.57]{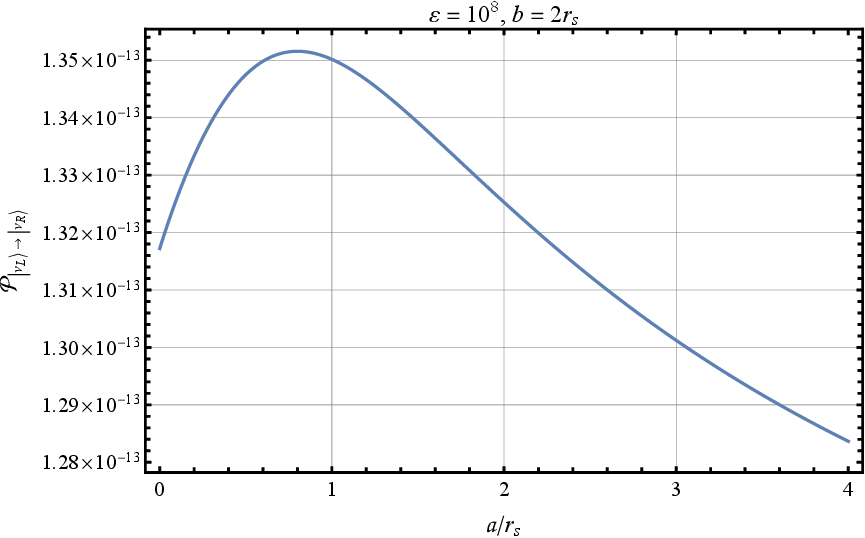}
    \caption{Spin-flip probability (given by Eq.\,(\ref{SpinProbabilityIntegralKerr})) for various values of the parameter $k=a/r_s$ for high-energy neutrinos ($\varepsilon=10^8$). The impact parameter $b$ is set to $2r_s$. This plot displays a peak in the probability for a specific value of the ratio $a/r_s$.}
    \label{fig:KerrHighEnergy}
\end{figure}
\subsubsection{Low-energy neutrinos}\label{RN Low e}
As with the previous spacetimes, low-energy neutrinos experience spin precession more than the high-energy neutrinos do, for the spin-flip probability $\mathcal{P}_{\ket{\nu_L}\rightarrow \ket{\nu_R}}$ one extracts from formula (\ref{SpinProbabilityIntegralKerr}) reaches unity. For this reason, instead of fixing the energy per unit mass $\varepsilon$, we are also going to vary the latter from its lowest possible value to $\varepsilon=3$. By doing so, we fix the black hole's rotation parameter $k=a/r_s$ to two different values, $k=0$ (corresponding to the Schwarzschild black hole) and $k=9/20$ (corresponding to a slowly rotating black hole). The numerical evaluation of integral (\ref{SpinProbabilityIntegralKerr}) then yields the two curves displayed in Fig.\,\ref{fig:KerrVaryEnergy} below. In both cases, the probability acquires a single peak where it reaches unity. Away from that peak, the probability decreases faster for lower values of $\varepsilon$ and relatively slower for larger values of $\varepsilon$. The peak values of the probability are given in the caption of that figure.
\begin{figure}[H]
    \centering
    \includegraphics[scale=0.57]{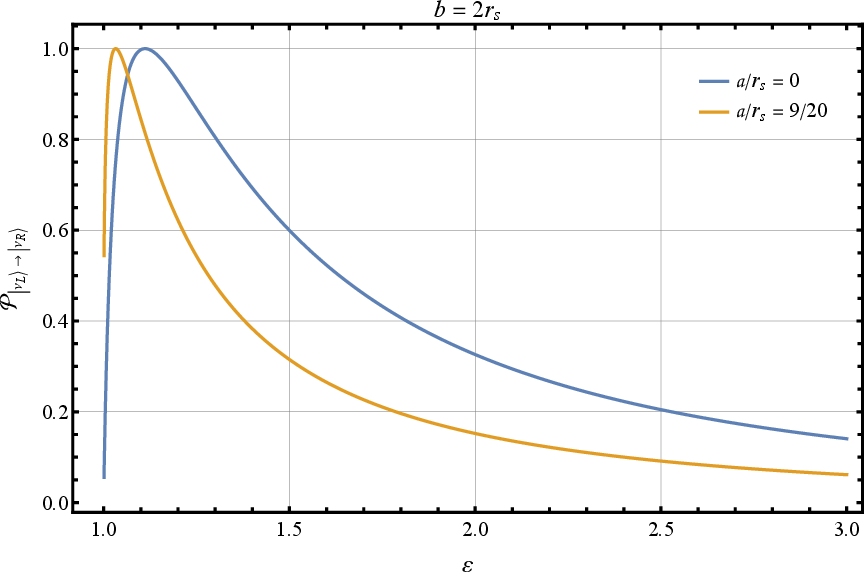}\\
        \caption{Variations of the spin-flip probability (given by Eq.\,(\ref{SpinProbabilityIntegralKerr})) for low-energy neutrinos. The impact parameter is fixed at $2r_s$ while the black hole's rotation parameter is chosen to take two different values, $a/r_s=0$ and $a/r_s=9/20$. The energy per unit mass $\varepsilon$ of the neutrinos is varied from its lowest possible value $\varepsilon=1$ to $\varepsilon=3$. The peaks in the probability reach unity and occur as follows: For $a/r_s=0$, the peak in the probability is achieved for $\varepsilon=1.112$ (like in the Schwarzschild spacetime). For $a/r_s=9/20$, the peak in the probability is achieved for $\varepsilon=1.032$.}
    \label{fig:KerrVaryEnergy}
\end{figure}

\subsection{Spin-flip probability in the wormhole spacetime}\label{SubSec:NeutrinosInWormhole}

Before proceeding with the computation of the spin-flip probability for the wormhole spacetime, we need to check again for which values of the angle $\theta$ does the magnitude of the angular velocity (\ref{AngularVelocitySWH}) we found for spin precession in this spacetime reach its maximum value. A long, but straightforward calculation of the derivative of the angular velocity's magnitude $\Omega$ with respect to $\theta$ using the two nonvanishing components given in Eq.\,(\ref{AngularVelocitySWH}) reveals that such a derivative vanishes for $\theta=\pi/2$. As such, we conclude that, just like the other spacetimes we dealt with above, spin precession does indeed reach its maximum angular speed when the particle propagates along the equatorial plane of the wormhole spacetime. 

Setting $\theta=\pi/2$ in the expressions (\ref{AngularVelocitySchwarMel}) yields a single nonvanishing component for the angular velocity vector that reads
\begin{equation}\label{EquatorAngularVelocityWormHole}
    \Omega_{\hat{2}}=\pm \frac{\varepsilon r \ell }{\sqrt{\alpha ^2+r^2} \left(\alpha ^2+r^2+\ell ^2\right)}.
    \end{equation}
Plugging this result into the general probability expression (\ref{GeneralProbabilityFormula}), and making the change of variables $x=\ell/r$ and setting $k=\alpha/r_s$ and $\mu=r_s/\ell$, we find the following probability for a left-handed neutrino to turn into a right-handed neutrino as it propagates and gets deflected in the spacetime:
\begin{equation}
\mathcal{P}(\ket{\nu_L}\rightarrow \ket{\nu_R})=\sin^2\left[ \int_{0}^{\sqrt{\varepsilon^2-1}} \mathcal{I}(x)\right],
\end{equation}\label{SpinProbabilityIntegralWH}
where the functions $\mathcal{I}(x)$ reads
\begin{align}
&\mathcal{I}(x)=\nonumber\\
&\frac{2\varepsilon \left(k^2 \mu ^2 x^2+1\right)^{\frac{1}{4}}\left(k^2\mu^2x^2+x^2+1\right)^{-1}}{ \sqrt{\varepsilon^2\left(k^2\mu ^2 x^2\!+\!1\right)^{\frac{3}{2}}\!-\!(k^2\mu^2x^2\!+\!x^2\!+\!1) (\sqrt{k^2 \mu ^2 x^2+1}\!-\!\mu x)}}.
\end{align}
The limits $0$ and  $\sqrt{\varepsilon^2-1}$ of integration over $x$ are obtained by, respectively, sending the radial coordinate $r$ to infinity and identifying the radial coordinate with the impact parameter $b$.

We consider separately in what follows high-energy and low-energy neutrinos. We start with the case of high-energy neutrinos by fixing $\varepsilon$ and plot the variations of the probability $\mathcal{P}_{\ket{\nu_L}\rightarrow \ket{\nu_R}}$ against the variations of the parameter $k$. We then proceed to study the variations of the probability for low-energy neutrinos by letting their energy per unit mass $\varepsilon$ vary from the lowest possible value $\varepsilon=1$ to $\varepsilon=3.5$ for four  different values of the regulator parameter $k=\alpha/r_s$ of the wormhole.

\subsubsection{High-energy neutrinos}\label{SWH High e}
We set $\varepsilon=10^8$, for which the parameter $\mu$ takes the value $\mu=2.5\times10^{-8}$. We fix the impact parameter to $b=2r_s$. We let the parameter $k=\alpha/r_s$ vary from $0$ to a value as high as $10^8$. The numerical evaluation of integral (\ref{SpinProbabilityIntegralWH}) for the different values of the parameter $k$ gives the result shown in Fig.\,\ref{fig:WormholeHighEnergy} below. The probability peaks at unity for $k=1.99\times10^8$ and decreases to zero away from that value of $k$; faster for smaller values of $\varepsilon$ and slower for larger values of $\varepsilon$.

Among all the spacetimes we considered up to now, this spinning wormhole is the second spacetime ---\,after the interior Schwarzschild spacetime and the Schwazschild-Melvin spacetime\,--- to induce large spin-flip probability even on high-energy neutrinos.
\begin{figure}[H]
    \centering
    \includegraphics[scale=0.57]{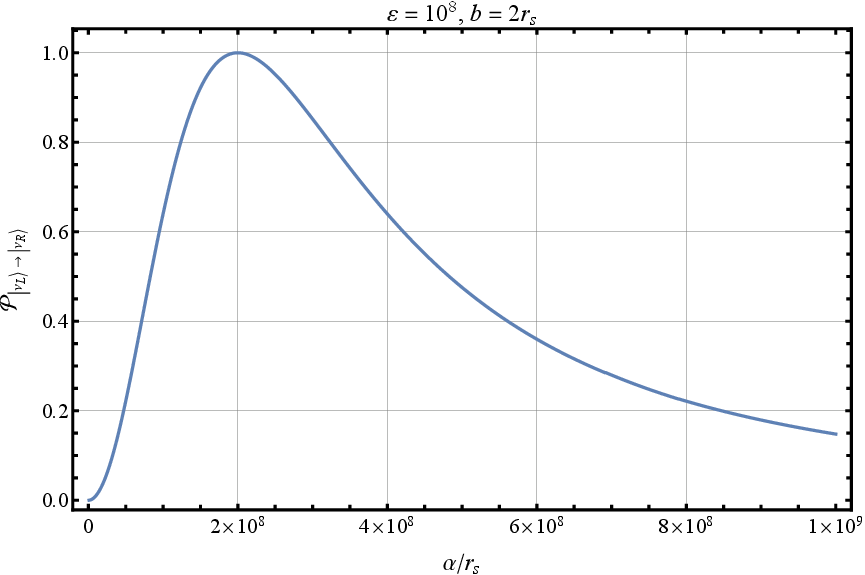}
    \caption{The variations of the spin-flip probability (given by Eq.\,(\ref{SpinProbabilityIntegralWH})) against the variations of the rotation parameter of the wormhole spacetime for high-energy neutrinos ($\varepsilon=10^8$) with an impact parameter $b=2r_s$. The peak value of the probability is one, and it is achieved for $\alpha/r_s=1.99\times10^8$.}
    \label{fig:WormholeHighEnergy}
\end{figure}

\begin{figure}[H]
    \centering
    \includegraphics[scale=0.57]{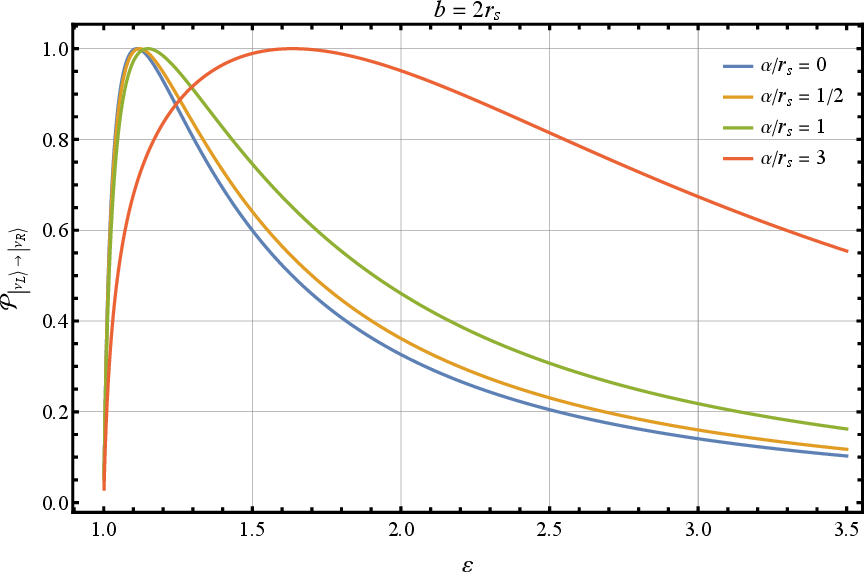}
    \caption{The four curves in this plot represent the variations of the spin-flip probability (given by Eq.\,(\ref{SpinProbabilityIntegralWH})) in terms of the variations of the energy per unit mass $\varepsilon$ of low-energy neutrinos. The impact parameter is chosen to be $b=2r_s$, but the wormhole's parameter $\alpha/r_s$ is assigned four different values: $0$, $1/2$, $1$ and $3$. The peaks in the probability all reach the value one and they are located as follows: For $\alpha/r_s=0$, the peak in the probability is reached for $\varepsilon=1.112$ (like in the Schwarzschild spacetime). For $\alpha/r_s=1/2$, the peak in the probability is reached for $\varepsilon=1.120$. For $\alpha/r_s=1$, the peak in the probability is reached for $\varepsilon=1.149$. For $\alpha/r_s=3$, the peak in the probability is reached for $\varepsilon=1.636$.}
    \label{fig:WormholelLowEnergy}
\end{figure}

\subsubsection{Low-energy neutrinos}\label{SWH Low e}
Just as for high-energy neutrinos, this spacetime also induces a large spin-flip probability for low-energy neutrinos. We vary here the energy per unit mass of the neutrinos and plot the variations of the probability for three different values of the wormhole's parameter $\alpha/r_s$: $\alpha/r_s = 0$, corresponding to the Schwarzschild spacetime; $\alpha/r_s = 1/2$ corresponding to a regular black hole; $\alpha/r_s = 1$, corresponding to a one-way wormhole with a null throat; $\alpha/r_s = 3$, corresponding to a two-way traversable wormhole \textit{\`a la} Morris-Thorne. We obtain the results displayed in Fig.\,\ref{fig:WormholelLowEnergy}. Spin-flip probability neatly peaks around values of $\varepsilon$ not very far from $\varepsilon=1$. However, the peak shifts faster towards larger values of $\varepsilon$ for a larger $\alpha/r_s$ parameter of the wormhole. Moreover, the probability decreases from these peaks slower when such a parameter is large and faster for smaller values of the parameter.
\subsection{The straight spinning cosmic string spacetime
}\label{SubSec:InStraightSpinningString
}
As the angular velocity of spin precession within the straight spinning cosmic string spacetime is different from all of the velocities found for the previous spacetimes, we also need first to verify here for which values of the angle $\theta$ does the magnitude of the angular velocity (\ref{AngularVelocitySSCS}) we found for spin precession in this spacetime reach its maximum value. A straightforward calculation of the derivative of the angular velocity's magnitude $\Omega$ with respect to $\theta$ using the two nonvanishing components given in Eq.\,(\ref{AngularVelocitySSCS}) shows that such a derivative vanishes again for $\theta=\pi/2$. As such, we conclude that, just like the other spacetimes we dealt with above, spin precession does indeed reach its maximum angular speed when the particle propagates along the equatorial plane of the straight spinning cosmic string. 

Setting $\theta=\pi/2$ in the expressions (\ref{AngularVelocitySSCS}) yields a single nonvanishing component for the angular velocity three-vector that reads
\begin{equation}\label{AngularVelocity2SSCS}
    \Omega_{\hat{2}}=\pm\frac{a \varepsilon (4 \varepsilon J+\ell )}{16 \varepsilon^2 J^2+8 \varepsilon \ell  J+a^2 r^2+\ell ^2}.
    \end{equation}
Plugging this result into the general probability expression (\ref{GeneralProbabilityFormula}), and making the change of variables $x=\ell/r$, we find the following probability integral (that can be evaluated analytically) for a left-handed neutrino to turn into a right-handed neutrino as it propagates and gets deflected in the spacetime:
\begin{align}
\mathcal{P}(\ket{\nu_L}\rightarrow \ket{\nu_R})=\sin^2\left[\int_{\infty}^{b}\mathcal{I}(r)\,{\rm d}r\right],
\end{align}
where
\begin{align}
    \mathcal{I}(r)&={2 a^2 \varepsilon r (4 \varepsilon J+\ell )}\left(a^2 r^2+16 \varepsilon^2 J^2+8 \varepsilon J \ell +\ell ^2\right)^{-1}\nonumber\\
    &\quad\times\Big{[}\varepsilon^2 \left(a^2 r^2-16 J^2\right)-a^2 r^2-8 \varepsilon J \ell -\ell ^2\Big{]}^{-1/2},\nonumber
\end{align}
leading to
\begin{align}\label{SpinProbabilitySSCS}
\mathcal{P}(\ket{\nu_L}\rightarrow \ket{\nu_R})&=\nonumber\\
&\frac{4 \varepsilon^2 \left[a^2 b^2 \left(\varepsilon^2\!-\!1\right) (4 \varepsilon J+\ell )^2\!-\!(4 \varepsilon J+\ell )^4\right]}{\left(\varepsilon^2-1\right)^2 \left[a^2 b^2+(4 \varepsilon J+\ell )^2\right]^2}.
\end{align}
The condition for having a positive probability is to have the sum inside the square root in the denominator in the first line of Eq.\,(\ref{SpinProbabilitySSCS}) positive. Such a condition is achieved for 
\begin{equation}
    a>1+\frac{4\varepsilon J}{b\sqrt{\varepsilon^2-1}}.
\end{equation}
This condition actually stems from requiring that the component $\beta$ of the four-velocity given by Eq.\,(\ref{alphabetagamma}) is also real. 
In what follows, we consider only low-energy neutrinos. The reason is that the probability formula given by Eq.\,(\ref{SpinProbabilitySSCS}) is an exact analytic expression that clearly shows that for high-energy neutrinos the probability approximates as $\mathcal{P}_{\ket{\nu_L}\rightarrow \ket{\nu_R}}=\mathcal{O}(1/\varepsilon^2)$. Therefore, high-energy neutrinos experience a spin-flip probability that decreases as the inverse square of their energies per unit mass, and hence remains insignificant. For low-energy neutrinos, we shall consider various scenarios according to the values of $a$ and $J/b$. 
\subsubsection{Low-energy neutrinos}\label{String Low e}
\begin{figure}[H]
    \centering
    \includegraphics[scale=0.57]{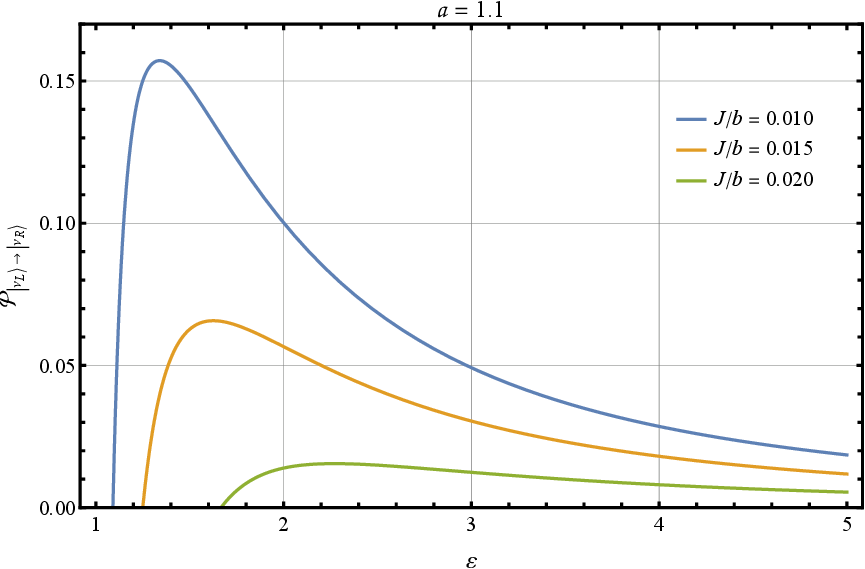}\\
        \caption{The variations of the spin-flip probability (\ref{SpinProbabilitySSCS}) for low-energy neutrinos in terms of the variations of the energy per unit mass $\varepsilon$ for three different values of the ratio $J/b$, but for the same value $a=1.1$. The two peaks displayed in this graph are located as follows: For $J/b=0.010$, the peak value of the probability is $0.15$ and it is located at $\varepsilon=1.34$. For $J/b=0.015$, the peak value of the probability is $0.13$ and it is located at $\varepsilon=1.62$. For $J/b=0.020$, the peak value of the probability is $0.08$ and it is located at $\varepsilon=2.27$.}
    \label{fig:CosmicStringIII}
\end{figure}

We first set $\varepsilon=1.1$ and vary the ratio $J/b$ for three different  values of the parameter $a$. Plotting the function (\ref{SpinProbabilitySSCS}) for those values of $a$ gives the results displayed in Fig.\,\ref{fig:CosmicStringI} below. The curves show only two peaks in the probability, both of value one; one for $a=2$ and another for $a=3$. Away from their peaks, the probabilities decrease rapidly to zero in all three cases.
\begin{figure}[H]
    \centering
    \includegraphics[scale=0.57]{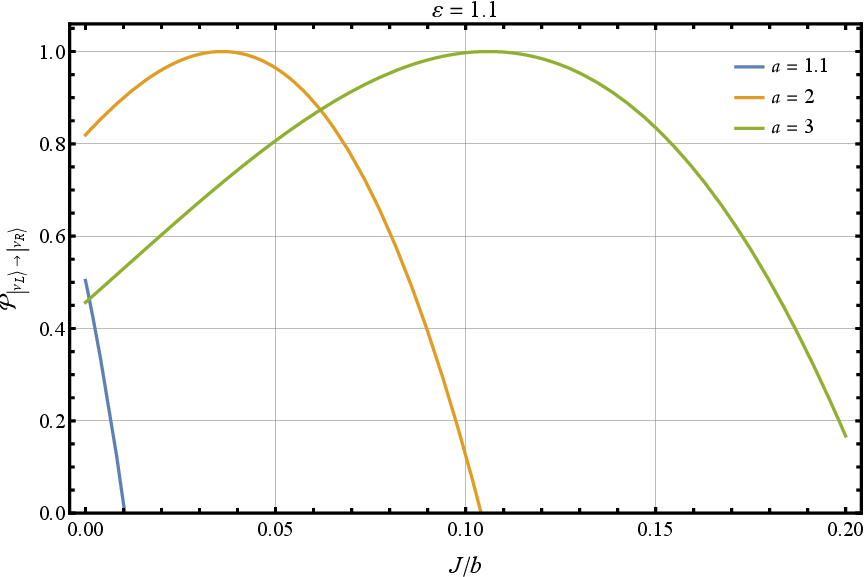}\\
        \caption{The variations of spin-flip probability (\ref{SpinProbabilitySSCS}) for low-energy neutrinos (with $\varepsilon=1.1$) in terms of the variations of the ratio $J/b$ for three different values of the parameter $a$. The two peaks displayed in this graph are located as follows: For $a=2$, the peak value of the probability is one and it is located at $J/b=0.035$. For $a=3$, the peak value of the probability is one and it is located at $J/b=0.106$.}
    \label{fig:CosmicStringI}
\end{figure}
We now set $\varepsilon=5$ and vary the ratio $J/b$ for the same three different values we previously chose for the parameter $a$. The plots in Fig.\,\ref{fig:CosmicStringII} below show that the spin-flip probability decreases in all three cases from a distinct maximum value and rapidly goes to zero for relatively small values of the ratio $J/b$. None of the three cases displays a change in its decreasing behavior. 
\begin{figure}[H]
    \centering
    \includegraphics[scale=0.57]{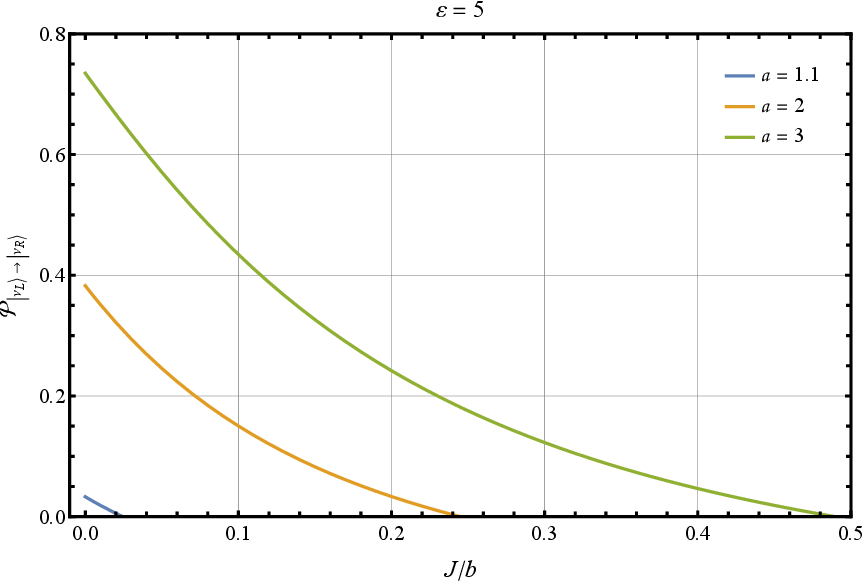}\\
        \caption{The variations of the spin-flip probability (\ref{SpinProbabilitySSCS}) for low-energy neutrinos (with $\varepsilon=5$) in terms of the variations of the ratio $J/b$ for three different values of the parameter $a$. All three cases display a rapid decrease in the probability from a certain maximum value for each case.}
    \label{fig:CosmicStringII}
\end{figure}
We now vary the energy per unit mass $\varepsilon$ over a range covering the lowest value $\varepsilon=1$ and going up to $\varepsilon=5$. We first do it by choosing the parameter $a=1.1$ and selecting three different values of the ratio $J/b$. For this case, the three curves display each a single peak value in the probability, without ever reaching unity though. All three curves displayed in Fig.\,\ref{fig:CosmicStringIII} show a rapid decrease towards zero for smaller values of $\varepsilon$ and a slow decrease towards zero for larger values of $\varepsilon$ away from their respective peaks.  The values of those peaks are given in the caption of the figure.
We finally vary now the energy per unit mass $\varepsilon$ over a range covering the lowest value $\varepsilon=1$ and going up to $\varepsilon=4$ by choosing $a=2$. Selecting again the same three distinct values of the ratio $J/b$, we plot the variations of the spin-flip probability against the variations of $\varepsilon$. The results are displayed in Fig.\,\ref{fig:CosmicStringIV} below. Remarkably, all three cases display two distinct maxima for the probability, separated in each case by a relative minimum. Away from their maxima, the probability curves decrease to zero rapidly for smaller values of $\varepsilon$ and slowly for large values of $\varepsilon$.  The values of those peaks and relative minima are given in the caption of the figure.
\begin{figure}[H]
    \centering
    \includegraphics[scale=0.57]{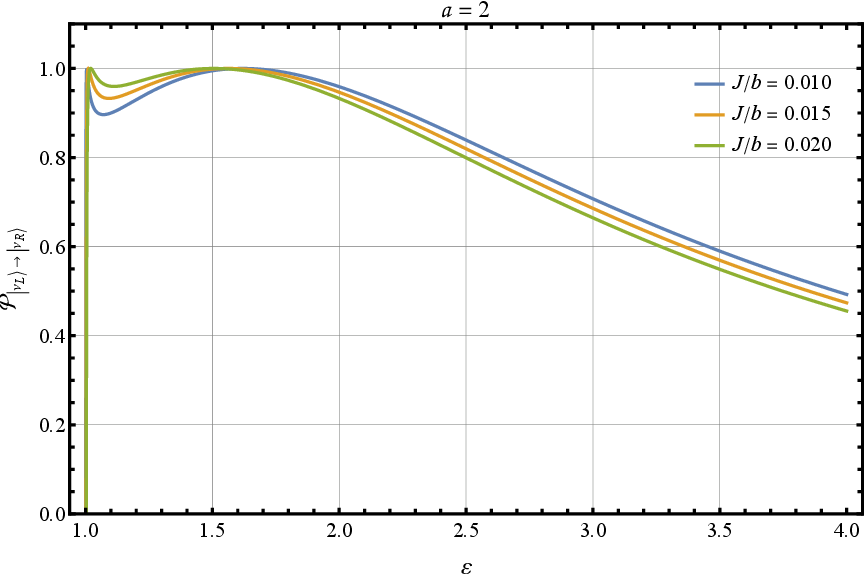}\\
        \caption{Variations of the spin-flip probability (\ref{SpinProbabilitySSCS}) for low-energy neutrinos in terms of the variations of the energy per unit mass $\varepsilon$ for three different values of the ratio $J/b$, but for the same value $a=2$. All the peaks in the probability have the value one. The two peaks displayed by each curve and the corresponding relative minima in between are located as follows: For $J/b=0.010$, the peak values of the probability are located at $\varepsilon=1.004$ and $\varepsilon=1.619$, respectively. The corresponding relative minimum is found for $\varepsilon=1.070$, for which the probability has the value $0.89$. For $J/b=0.015$, the peak values of the probability are located at $\varepsilon=1.011$ and $\varepsilon=1.562$, respectively. The corresponding relative minimum is found for $\varepsilon=1.092$, for which the probability has the value $0.93$. For $J/b=0.020$, the peak values of the probability are located at $\varepsilon=1.020$ and $\varepsilon=1.504$, respectively. The corresponding relative minimum is found for $\varepsilon=1.112$, for which the probability has the value $0.95$.}
    \label{fig:CosmicStringIV}
\end{figure}



\section{Summary and conclusion}\label{Sec:Conclusion}
We have studied in this paper the spin precession of neutral spinning particles freely propagating within various curved spacetimes. We did so by first deriving the formula that describes the spin precession of neutral particles moving along geodesics in a general stationary and axisymmetric spacetime. Our result is then applied to study spin precession of neutral spinning particles moving within thirteen different spacetimes. Those spacetimes have been selected among the most frequently encountered in the literature. We derived for each of those spacetimes the general expression of the angular velocity three-vector of spin precession.

The angular velocity three-vector obtained for spin precession in all the spacetimes considered here had only, at most, two nonvanishing components, except in the Weyl spacetime. The angular velocity of spin precession within the latter spacetime is found to have all of its three components nonvanishing. On the other hand, the majority of the spacetimes we considered here gave us an angular velocity for spin precession that vanishes for particles of zero orbital angular momentum, except in the Weyl spacetime, the Kerr spacetime and the straight spinning cosmic string spacetime. This means that a spinning particle in these three spacetimes would undergo spin precession even if the particle had initially a zero angular momentum. 

We then applied all those expressions of the angular velocity of spin precession to study helicity flip of neutrinos freely propagating in each one of those selected spacetimes. We used the general spin-flip probability integral together with the various expressions we obtained for the angular velocity of spin precession to derive the probability formula for a left-handed neutrino to turn into a right-handed neutrino when it freely propagates in each one of those spacetimes that do induce spin precession. We used those results to study the variations of the probability in terms of the neutrinos' energy per unit mass as well as the various parameters of each spacetime considered. 

A very interesting result among the ones we obtained is that while in some of these spacetimes spin-flip probability remains insignificant for low-energy neutrinos, a few of the spacetimes considered here do allow spin-flip probability to be as high as unity even for high-energy neutrinos. Those spacetimes are the Melvin-Bonnor spacetime, the interior Schwarzschild solution, the Schwarzschild-Melvin spacetime and the wormhole spacetime. Among these four spacetimes, the interior Schwarzschild solution, albeit more idealistic than real, is probably the most interesting case as it is the spacetime that approximately describes the gravitational field inside static astrophysical gravitational sources. Our results for this case show indeed that, depending on the size of the gravitational source, and for a specific value of the impact parameter of the high-energy neutrinos, the latter acquire a probability to flip their spin that could be as large as one. On the other hand, all the spacetimes that do induce spin precession were found to lead to a significant probability for spin flip when the neutrinos propagate at low energies. 

A remarkable result we found concerning spin-flip probability is that in some spacetimes, low-energy neutrinos acquire a spin-flip probability that displays two distinct peaks, each of value unity, but located at different values of the energy per unit mass $\varepsilon$ of the neutrinos. Those spacetimes are the Kiselev spacetime, the Schwarzschild-Melvin spacetime and the straight spinning cosmic string spacetime. In between the consecutive peaks, the probability dips to a relative minimum value. Mathematically, this comes about due to the presence of the sine function in all the probability formulas. Those spacetimes that do give a highly oscillatory argument for the sine function make the probability reach twice its maximum value of one. Physically, this comes about due to the fact that these three spacetimes are each a two-parameter spacetime. As a consequence, the single resonance that we obtained for the interior Schwarzschild spacetime is manifested twice in these three spacetimes, and its location depends on the energy per unit mass of the deflected particle as well as on the values those two parameters take. 
Concerning the absence of two peaks in the other two-parameter spacetimes, we notice that only those two-parameter spacetimes that possess two horizons and only when the particle is moving in between those two horizons that two probability peaks arise. Spacetimes within which the particle moves outside both horizons do not induce two peaks since in those cases the particle sees only one horizon and propagates as if it were inside a one-parameter spacetime.

The only spacetime that has two parameters and no horizons and, yet, induces two peaks in the probability is the straight spinning cosmic string. We can understand the origin of the two peaks for this case by noticing that the deficit angle created by the cosmic string depends on two parameters $a$ and $J$, and that the value of the angular deficit seen by the particle depends on the location $r$ of the particle at deflection. The latter location is decided by the impact parameter $b$. On the other hand, the graphs in Fig.\,\ref{fig:CosmicStringIV} are obtained not by fixing both $J$ and $b$, but by fixing only the ratio $J/b$. This means that the resonance occurs for two possible values of $\varepsilon$ as the impact parameter $b$ is allowed accordingly to take different values as long as the ratio $J/b$ remains the same. This is what exposes the effect of the two parameters in this spacetime: it is the variation of $J$ and $b$ at the same time, which both make the particle see the same deficit angle for two different values of $\varepsilon$ but for a fixed $a$. The deficit angle together with the spinning parameter $J$ of the string are what decide how much the particle precesses.
\section*{Acknowledgments}
This work was supported by the Natural Sciences and Engineering Research Council of Canada (NSERC) Discovery Grant No. RGPIN-2017-05388; and by the Fonds de Recherche du Québec - Nature et Technologies (FRQNT).

\appendix
\section{The comoving vierbeins and the spin connection for the general metric (\ref{GeneralMetric}) and four-velocity (\ref{General4Velocity})}\label{Sec:App}
We display here the four comoving vierbeins we extract from the general metric (\ref{GeneralMetric}). Using the four-velocity (\ref{General4Velocity}), the comoving vierbeins read,
\begin{align}\label{AppOrbitSchwarVierbeins}
    e_{\hat 0}^\mu&=\left(\frac{\alpha}{\gamma},\pm\frac{\beta}{\gamma},0,\frac{\lambda}{\gamma}\right),\nonumber\\
e_{\hat 1}^\mu&=\left(\frac{\mathcal{R}\beta}{\sqrt{\Xi}},\pm\frac{\Lambda\alpha+\Psi\lambda}{\sqrt{\Xi}},0,0\right),\nonumber\\
e_{\hat2}^\mu&=\left(0,0,\pm\frac{1}{\sqrt{\Theta}},0\right),\nonumber\\
e_{\hat 3}^\mu&= \Bigg(\frac{\left(\lambda\Phi-\Psi\alpha\right)\left(\Lambda\alpha+\Psi\lambda\right)+\mathcal{R}\beta^2\Psi}{\sqrt{\Delta}},\nonumber\\
&\qquad\pm\frac{\lambda\beta\left(\Lambda\Phi+\Psi^2\right)}{\sqrt{\Delta}},0,\frac{\left(\alpha\Lambda+\Psi\lambda\right)^2-\mathcal{R}\beta^2\Lambda}{\sqrt{\Delta}}\Bigg),
\end{align}
where
\begin{align}\label{Delta}
&\Xi=\mathcal{R}(\Lambda\alpha+\Psi\lambda)^2-\Lambda\mathcal{R}^2\beta^2,\nonumber\\
&\Delta=\left(\Lambda\Phi+\Psi^2\right)\!\Big{\{}\mathcal{R}\beta^2\lambda^2\left(\Lambda\Phi+\Psi^2\right)+\mathcal{R}^2\beta^4\Lambda\nonumber\\
&+\left(\Lambda\alpha+\Psi\lambda\right)^2[\alpha\left(\Lambda\alpha+\Psi\lambda\right)+\lambda\left(\Psi\alpha-\lambda\Phi\right)-2\mathcal{R}\beta^2]\Big{\}}.
\end{align}
The inverse vierbeins $e^{\hat a}_\mu$ take then the following form:
\begin{align}\label{AppInverseOrbitSchwarVierbeins}
    e^{\hat 0}_\mu&=\left(\frac{\Lambda\alpha+\Psi\lambda}{\gamma},\,\mp\frac{\mathcal{R}\beta}{\gamma},\,0,\,\frac{\Psi\alpha-\Phi\lambda}{\gamma}\right),\nonumber\\ 
    e^{\hat 1}_\mu&=\left(-\frac{\Lambda\mathcal{R}\beta}{\sqrt{\Xi}},\pm\frac{\mathcal{R}(\Lambda\alpha+\Psi\lambda)}{\sqrt{\Xi}},0,-\frac{\Psi\mathcal{R}\beta}{\sqrt{\Xi}}\right),\nonumber\\ 
    e^{\hat 2}_\mu&=\left(0,0,\pm \sqrt{\Theta},0\right),\nonumber\\
    e^{\hat 3}_\mu&=\Bigg(-\frac{\lambda\left(\Lambda\alpha+\Psi\lambda\right)\left(\Lambda\Phi+\Psi^2\right)}{\sqrt{\Delta}},\pm\frac{\mathcal{R}\beta\lambda\left(\Lambda\Phi+\Psi^2\right)}{\sqrt{\Delta}},\nonumber\\
&\qquad0,\frac{\left[\alpha\left(\Lambda\alpha+\Psi\lambda\right)-\mathcal{R}\beta^2\right]\left(\Lambda\Phi+\Psi^2\right)}{\sqrt{\Delta}}\Bigg).
\end{align}
The inverse metric of metric (\ref{GeneralMetric}) being
\begin{align}\label{InverseMetric}
g^{\mu\nu}=\begin{pmatrix}
    -\Phi(\Lambda\Phi+\Psi^2)^{-1} & \quad 0 & \quad 0 & -\Psi(\Lambda\Phi+\Psi^2)^{-1}\\
    0 & \mathcal{R}^{-1} & 0 & 0 \\
    0 & 0 & \Theta^{-1} & 0 \\
    -\Psi(\Lambda\Phi+\Psi^2)^{-1} & \quad 0 & \quad 0 & \Lambda(\Lambda\Phi+\Psi^2)^{-1}\\
\end{pmatrix},
\end{align}
the nonzero Christoffel symbols we extract from this general metric are the following:
\begin{align}
\Gamma_{01}^0&=\frac{\Psi\Psi_{,r}+\Phi\Lambda_{,r}}{2(\Psi^2+\Phi\Lambda)},\qquad\Gamma_{00}^1=\frac{\Lambda_{,r}}{2\mathcal{R}},\quad\qquad \Gamma^{2}_{00}=\frac{\Lambda_{,\theta} }{2 \Theta },\nonumber\\
\Gamma_{13}^0&=\frac{\Phi\Psi_{,r}-\Psi\Phi_{,r}}{2(\Psi^2+\Phi\Lambda)},\qquad\Gamma_{11}^1=\frac{\mathcal{R}_{,r}}{2\mathcal{R}},\quad\qquad \Gamma^{2}_{03}=\frac{\Psi_{,\theta} }{2 \Theta },\nonumber\\
\Gamma^{0}_{02}&=\frac{\Phi\Lambda_{,\theta} +\Psi\Psi_{,\theta} }{2 \left(\Psi^2+\Lambda  \Phi\right)},\quad\quad\Gamma^{1}_{12}=\frac{\mathcal{R}_{,\theta}}{2 \mathcal{R}},\;\;\;\qquad \Gamma^{2}_{11}=-\frac{\mathcal{R}_{,\theta}}{2 \Theta },\nonumber\\
\Gamma^{0}_{23}&=\frac{\Phi\Psi_{,\theta} -\Psi  \Phi_{,\theta} }{2 \left(\Psi ^2+\Lambda  \Phi\right)},\qquad\Gamma_{22}^1=-\frac{\Theta_{,r}}{2\mathcal{R}},\!\!\;\qquad \Gamma_{12}^2=\frac{\Theta_{,r}}{2\Theta},\nonumber\\
\Gamma_{01}^3&=\frac{\Psi\Lambda_{,r}-\Lambda\Psi_{,r}}{2(\Psi^2+\Phi\Lambda)},\;\qquad\Gamma_{33}^1=-\frac{\Phi_{,r}}{2\mathcal{R}},\!\!\,\;\;\qquad  \Gamma^{2}_{22}=\frac{\Theta_{,\theta} }{2 \Theta },\nonumber\\
\Gamma^{3}_{02}&=\frac{\Psi  \Lambda_{,\theta}-\Lambda\Psi_{,\theta} }{2 \left(\Lambda  \Phi +\Psi ^2\right)},\!\!\!\qquad\qquad\qquad\qquad\qquad\Gamma^{2}_{33}=-\frac{\Phi_{,\theta} }{2 \Theta },\nonumber\\
\Gamma_{13}^3&=\frac{\Psi\Psi_{,r}+\Lambda\Phi_{,r}}{2(\Psi^2+\Phi\Lambda)},\nonumber\\
\Gamma^{3}_{23}&=\frac{\Lambda\Phi_{,\theta} +\Psi  \Psi_{,\theta} }{2 \left(\Psi ^2+\Lambda  \Phi\right)}.
\end{align}
Here, a partial derivative of a function $f$ with respect to a coordinate $x^\mu$ is denoted by $f_{,\mu}$. From these expressions, we compute the relevant nonzero coefficients of the spin connection to be:
\begin{align}\label{AppOrbitSchwarSpinConnect}
    \omega^{\hat{1}\hat{2}}_0&= \mp \frac{\beta  \mathcal{R} \Lambda_{,\theta} }{2 \sqrt{\Theta  \Xi }},\nonumber\\
    \omega^{\hat{1}\hat{2}}_1 &= \frac{\mathcal{R}_{,\theta} (\alpha  \Lambda +\lambda  \Psi )}{2 \sqrt{\Theta  \Xi }},\nonumber\\
    \omega^{\hat{1}\hat{2}}_2 &=-\frac{ \Theta_{,r}  (\alpha  \Lambda +\lambda  \Psi )}{2 \sqrt{\Theta  \Xi }},\nonumber\\
    \omega^{\hat{1}\hat{2}}_3 &=\mp \frac{\beta  \mathcal{R} \Psi_{,\theta} }{2 \sqrt{\Theta  \Xi }},\nonumber\\
\omega^{\hat{2}\hat{3}}_0&=\pm\frac{1}{2 \sqrt{\Delta\Theta }}\Big\{\Psi_{,\theta}\left[(\alpha  \Lambda +\lambda  \Psi )^2-\beta ^2 \Lambda  \mathcal{R}\right]\nonumber\\
&+\Lambda_{,\theta}\left[\alpha  \lambda  \Lambda  \Phi\!-\!\alpha ^2\Lambda\Psi \!-\!\alpha  \lambda  \Psi ^2\!+\!\lambda ^2 \Phi  \Psi \!+\!\beta ^2 \Psi  \mathcal{R}\right]
    \Big\},\nonumber\\
    \omega^{\hat{2}\hat{3}}_1 &= -\frac{\beta  \lambda}{2 \sqrt{\Delta  \Theta }}  \mathcal{R}_{,\theta} \left(\Psi ^2+\Lambda  \Phi \right),\nonumber\\
    \omega^{\hat{2}\hat{3}}_2& = \frac{\beta  \lambda}{2\sqrt{\Delta\Theta}}\Theta_{,r}  \left(\Psi^2+\Lambda\Phi\right),\nonumber\\
    \omega^{\hat{2}\hat{3}}_3 &= \pm\frac{1}{2 \sqrt{\Delta\Theta}} \Big\{
    \beta ^2 \mathcal{R} \left[\Lambda  \Phi_{,\theta} +\Psi\Psi_{,\theta} \right]\nonumber\\
    &-(\alpha  \Lambda +\lambda  \Psi ) \left[\Phi_{,\theta}  (\alpha  \Lambda +\lambda\Psi)+\Psi_{,\theta}  (\alpha  \Psi -\lambda\Phi)\right]
    \Big\},\nonumber\\
    \omega^{\hat{1}\hat{3}}_0&=\pm\frac{\left(\Psi ^2+\Lambda  \Phi\right)^2}{2 \sqrt{\Delta  \Xi }}
    (\varepsilon\Psi_{,r} +\ell \Lambda_{,r} )\nonumber\\
    &\quad\times\left[\Psi ^2 \left(\varepsilon^2+\Lambda \right)+2 \varepsilon \Lambda  \ell \Psi +\Lambda ^2 \left(\ell^2+\Phi \right)\right]
    ,\nonumber\\
\omega^{\hat{1}\hat{3}}_1 &= -\frac{\left(\Psi ^2+\Lambda  \Phi\right)^2}{2 \beta  \sqrt{\Delta  \Xi }}{\Big[}\Psi_{,r}  \Big\{\varepsilon^4 \Phi  \Psi ^2+\varepsilon^2 \left(\Psi ^4-\Lambda ^2 \Phi ^2\right)\nonumber\\
&\quad+2 \varepsilon \Lambda ^2 \ell^3 \Psi +\Lambda  \left(\Lambda  \Phi +\Psi ^2\right)^2+\Lambda ^3 \ell^4\nonumber\\
    &\quad+\Lambda\ell^2 \left(\varepsilon^2+2 \Lambda \right) \left(\Lambda  \Phi +\Psi ^2\right)\nonumber\\
    &\quad+2 \varepsilon \Lambda  \ell \Psi  \left[\varepsilon^2 \Phi +2 \left(\Lambda  \Phi +\Psi ^2\right)\right]\Big\}\nonumber\\
    &\quad+\Lambda_{,r} \Big\{\ell^2 \varepsilon^2\left(\Psi ^3-2 \Lambda \Phi \Psi\right)-\varepsilon^3\ell \Phi  \Psi ^2\nonumber\\
    &\quad-2 \Phi  \Psi\varepsilon^2  \left(\Lambda \Phi +\Psi ^2\right)+\Psi  \left(\Lambda  \Phi +\Lambda  \ell^2+\Psi ^2\right)^2\nonumber\\
    &\quad+\varepsilon\ell \left[2 \Lambda  \Phi  \Psi ^2+\Lambda  \ell^2 \left(2 \Psi ^2-\Lambda  \Phi \right)+3 \Psi ^4-\Lambda ^2 \Phi ^2\right] \Big\}\nonumber\\
    &\quad+\varepsilon\Phi_{,r} (\varepsilon \Psi +\Lambda  \ell)^3{\Big]},\nonumber\\
    \omega^{\hat{1}\hat{3}}_2 &=\omega^{\hat{1}\hat{3}}_3= -\frac{\beta  \left(\Psi ^2+\Lambda\Phi\right)}{2 \sqrt{\Delta\Xi }}\nonumber\\
    &\times\Big\{\mathcal{R} \Psi_{,\theta} \left[\Psi ^2 \left(\varepsilon^2+\Lambda \right)+2 \varepsilon \Lambda  \ell \Psi +\Lambda ^2 \left(\ell^2+\Phi \right)\right]\nonumber\\
&+\mathcal{R} \Lambda_{,\theta}\left[\varepsilon\ell \left(\Psi^2\!-\!\Lambda\Phi\right)+\Lambda  \Psi\left(\ell^2\!+\!\Phi\right)\!+\!\Psi ^3-\varepsilon^2\Phi\Psi\right]\nonumber\\
&+\varepsilon \mathcal{R}_{,\theta}\left(\Psi ^2+\Lambda\Phi\right) (\varepsilon\Psi+\Lambda\ell)\Big\},\nonumber\\
    \omega^{\hat{1}\hat{3}}_4 &=\pm\frac{\left(\Psi ^2+\Lambda\Phi\right)^2}{2 \sqrt{\Delta\Xi }}(\ell\Psi_{,r}-\varepsilon\Phi_{,r})\nonumber\\
    &\quad\times\left[\Psi ^2 \left(\varepsilon^2+\Lambda\right)+2 \varepsilon \Lambda\ell\Psi +\Lambda ^2 \left(\ell^2+\Phi \right)\right],\nonumber\\
\end{align}


\section{The components of the angular velocity vector of spin precession obtained within the Kerr spacetime}\label{Sec:AppB}
The three components of the angular velocity vector are obtained by plugging the metric components (\ref{KerrMetric}) into the general expressions (\ref{GeneralAngularVelocity}). We the following expressions:
\begin{equation}\label{AppBOmega1}
    \Omega_{\hat{1}}=\frac{P}{Q},
\end{equation}
where
\begin{align}
    P&=\mp A \rho ^2 \cos \theta \Bigg\{A \varepsilon X^2 \ell  \sin ^2\theta \left[a A (a \varepsilon-\ell )-Y \left(a^2+r^2\right)\right]\nonumber\\
    &+\Bigg(\frac{4aZ}{\rho^2}
     \sin ^2\theta \left(a^2-A+r^2\right) \left[Y \left(a^2+r^2\right)+a A (\ell -a e)\right]\nonumber\\
     &\quad-\frac{ZX}{\rho ^4}\left(2 a^2 \sin ^2\theta+\rho ^2\right)\nonumber\\
     &\quad\times\left(a A \varepsilon \cos 2\theta-a A \varepsilon+2 a Y \sin ^2\theta+2 A \ell \right)\Bigg)\nonumber\\
     &\times\Bigg(A \sin ^2\theta \left[\left(a^2+r^2\right)^2-a^2 \left(a^2+2 r^2\right) \sin ^2\theta\right]\nonumber\\
     &\quad+A \rho ^2 \left[a \varepsilon \sin ^2\theta (a \varepsilon-2 \ell )+\ell ^2\right]-\rho ^2 Y^2 \sin ^2\theta\nonumber\\
     &\quad-a^2 A^2 \sin ^2\theta \cos ^2\theta\Bigg)
    \Bigg\},
    \end{align}
and
\begin{align}
    Q&=4 T \sin^3\theta A^2Z^2 \Bigg\{
    A \sin ^2\theta \Big(a^4 A-a^2 A^2 \cos ^2\theta+2 a^2 A r^2\nonumber\\
    &\quad+a \rho ^2 V Y+A r^4\Big)-a^4 A^2 \sin ^6\theta\nonumber\\
    &\quad-\frac{1}{2} a^2 \sin ^4\theta \Big[a^4 A+a^2 A \left(a^2+2 r^2\right) \cos 2\theta\nonumber\\
    &\quad+2 a^2 A r^2+4 A^2 r^2+2 A r^4-2 \rho ^2 Y^2\Big]+\frac{1}{4} A^2 \rho ^2 V^2
    \Bigg\}^{1/2}.
\end{align}

\begin{equation}\label{AppBOmega2}
    \Omega_{\hat{2}}=\frac{U}{W},
\end{equation}
where
\begin{align}
    U=&\mp A^2 Z^2 \sin \theta\Bigg\{
    2 a^2 A^3 r \cos ^2\theta \left(\varepsilon V-2 a \sin ^2\theta\right)\nonumber\\
    &+ 4 a^2\rho ^2 r Y^2 \sin ^4\theta\nonumber\\
    &+ a \rho ^2 Y \sin ^2\theta A' \Big[2 a^4 \varepsilon-a^3 \ell\nonumber\\
    &+a^2 \cos 2\theta \left(2 a^2 \varepsilon-a \ell +3 \varepsilon r^2\right)\nonumber\\
    &+5 a^2 \varepsilon r^2-2 a r^2 \ell +4 \varepsilon r^4\Big]\nonumber\\
    &+a^3FA^2 A' \sin ^2\theta \cos ^2\theta-a^2 \varepsilon \rho ^2 VA^2 A' \cos ^2\theta\nonumber\\
    &-4 a r A^2\sin ^2\theta \left[a^4 \sin ^4\theta+2 a^2 r^2 \sin ^2\theta-\left(a^2+r^2\right)^2\right]\nonumber\\
    &+4 \rho ^2 r A^2\Big[a^3 \varepsilon^2 \sin ^4\theta+\ell  \left(2 a^2 \varepsilon+a \ell +2 \varepsilon r^2\right)\nonumber\\
    &-2 a \varepsilon \sin ^2\theta \left(2 a \ell +\varepsilon r^2\right)\Big]\nonumber\\
    &-\varepsilon r A^2 \Big[a^5 \varepsilon-\tfrac{5}{2}a^4 \ell+\tfrac{3}{2}a^3 \varepsilon r^2\nonumber\\
    &-2 a \cos 2\theta \left[a \ell  \left(a^2+2 r^2\right)+\varepsilon r^4\right]-4 a^2 r^2 \ell\Big]\nonumber\\
    &+\varepsilon rA^2\Big[\tfrac{1}{2}a^3\cos 4\theta \left(2 a^2 \varepsilon\!-\!a \ell\!+\!3 \varepsilon r^2\right)\!+\!2 a \varepsilon r^4\!-\!48 r^4 \ell \Big]\nonumber\\
    &-\tfrac{1}{2}a A \sin ^2\theta \left[a^4+a^2 \left(a^2+2 r^2\right) \cos 2\theta+2 a^2 r^2+2 r^4\right]\nonumber\\
    &\times\left[4 r \left(a^2 \sin ^2\theta+\varepsilon Y\right)+F A'\right]\nonumber\\
    &-8 a^3\rho^2 r A \sin ^2\theta \left[a \varepsilon \sin ^2\theta (a \varepsilon-2 \ell )+\ell ^2\right]\nonumber\\
    &\quad+2a\rho^2 AA' \Big[a^2 \varepsilon \sin ^4\theta \left(3 a^2 \varepsilon-4 a \ell +2 \varepsilon r^2\right)\nonumber\\
    &\quad-\ell ^2 \left(a^2+r^2\right)\Big]-2a^2\rho^2AA' \sin ^2\theta\nonumber\\
    &\quad\quad\quad\times\left(3 a^3 \varepsilon^2-4 a^2 \varepsilon \ell +3 a \varepsilon^2 r^2-a \ell ^2-2 \varepsilon r^2 \ell \right)
    \Bigg\}.
\end{align}

\begin{align}
    W&= 4 \rho ^2 A^2Z^{\frac{5}{2}}\left(a^2-A+r^2\right)^2\nonumber\\
    &\times\Bigg\{A \sin ^2\theta \left(a^4 A\!-\!a^2 A^2 \cos ^2\theta\!+\!2 a^2 A r^2\!+\!a \rho ^2 V Y\!+\!A r^4\right)\nonumber\\
    &-a^4 A^2 \sin ^6\theta+2 a^2 A r^2+4 A^2 r^2+\frac{1}{4} A^2 \rho ^2 V^2
    \nonumber\\
&-\!\frac{a^2}{2}\!\sin^4\!\theta\!\left[a^4 A\!\!+\!\!a^2 A \left(a^2\!\!+\!\!2 r^2\right)\!\cos2\theta\!+\!2 A r^4\!\!-\!2 \rho ^2 Y^2\right]\!\!\!
    \Bigg\}^{\!\!\!1/2}\nonumber\\
    &\times\Bigg\{
    a^2 \varepsilon^2 \sin ^4\theta\nonumber\\
    &+a \sin ^2\theta \left(a^2-A+r^2\right) \left(\frac{a \sin ^2\theta}{\rho ^2}+\frac{2 \varepsilon \ell}{a^2-A+r^2} \right)\nonumber\\
    &+\!\left(A\!-\!a^2\!\sin^2\theta\right)^{\!2}\! \left[\frac{\left(a^2\!+\!r^2\right)^2\sin ^2\theta\!-\!a^2 A\sin ^2\theta}{\rho ^2}\!+\!\ell^2\right]\!\!
    \Bigg\}^{\!\!1/2}\!\!\!,
\end{align}

\begin{equation}
    \Omega _{\hat{3}} = 0.
\end{equation}
In all the above expressions, a prime stands for a derivative with respect to $r$ and we have set
\begin{align}
    X&=a^4-2 a^2(r^2-r_sr+a^2) \cos ^2\theta+a^2 \left(a^2+2 r^2\right) \cos 2\theta\nonumber\\
    &\quad+2 a^2 r^2+2 r^4,\nonumber\\
    Y&=\varepsilon(a^2+r^2)-a \ell,\nonumber\\
    Z&=\left(a^2+r^2\right)^2-a^2(r^2-r_sr+a^2) \cos ^2\theta\nonumber\\
    &\quad-a^2 \left(a^2+2 r^2\right) \sin ^2\theta,\nonumber\\
    V&=2( \ell-a\varepsilon\sin^2\theta),\nonumber\\
    F&=2(r^2-a^2\sin^2\theta).
\end{align}

\end{document}